# Coupled plastic strain- and stress-induced phase transformations and microstructure evolution in Fe-7%Mn alloy in a dynamic rotational diamond anvil cell

Sorb Yesudhas[1*°], Mrinmay Sahu[1], Valery I. Levitas[1,2*°], Dean Smith[3], Aniket Singh[1], Stanislav Sinogeikin[4], and Jeffrey T. Lloyd[5]

[1]Department of Aerospace Engineering, Iowa State University, Ames, Iowa 50011, USA

[2]Department of Mechanical Engineering, Iowa State University, Ames, Iowa 50011, USA

[3]HPCAT, X-ray Science Division, Argonne National Laboratory, Argonne, Illinois 60439, USA

[4]DAC Tools, LLC, Naperville, Illinois 60565, USA

[5]Armor Mechanisms Branch, DEVCOM Army Research Laboratory, Aberdeen Proving Ground, Maryland 21005, USA

*Corresponding authors. Email: sorbya@iastate.edu, vlevitas@iastate.edu

°These authors contributed equally to this work.

**Abstract**

Studies of severe plastic deformation (SPD), microstructure evolution, and plastic strain-induced phase transformations (PTs) are crucial for various fundamental and applied disciplines and phenomena. However, they are performed either quasi-statically or at low strain rates and pressures. Here, in situ experiments in a dynamic rotational diamond anvil cell (dRDAC) on SPD and BCC↔HCP PTs at pressures up to 27.6 GPa, rotation rates up to 1,500 RPM, and strain rates up to 2,299/s are performed, considering Fe-7%Mn alloy as an example. Strong plastic straining leads to a unique mechanism and kinetics with simultaneous direct and reverse PTs, which has not been studied for any material. For quasi-static loading, we determine the kinetic parameters for strain-induced direct-reverse PTs and the stationary volume fraction $c$ versus pressure. During torsion at 1,000 and 1,500 RPM, $c$ does not change. After torsion stops, 51 minutes later, it increases by 31% at 1,000 RPM and $c \to 0$, 7 minutes later at 1,500 RPM. These findings contradict the general wisdom that strain-induced PTs occur only during straining and are governed by strain, independent of time; they reveal an alternative mechanism. It is revealed that the crystallite size of

~28(6) nm, microstrain ~0.0035(8), and dislocation density ~1.3(6) × $10^{15}/m^2$ in the HCP phase are steady during static compression and dynamic torsion, during and after the PT, and after torsion. These parameters are independent of pressure, plastic strain tensor, its path, strain rates, and *c*. The results obtained open fundamental research on combined strain- and stress-induced PTs and microstructure evolution under dynamic SPD and high pressure, with various important applications.

## Introduction

PTs that occur during SPD under high pressure play a crucial role in physical and material experiments[1-3], the synthesis of high-pressure phases[1-5] and nanostructured materials[6,7], bulk and surface processing (cutting, polishing, and lapping) of materials[8,9], friction, military applications[10], and the interpretation of geological behavior and deep-focus earthquake phenomena[11,12]. Plastic deformation reduces PT by one to two orders of magnitude compared with hydrostatic loading, enabling the formation and stabilization of novel high-pressure phases and often revealing entirely new transformation pathways[1-5,13-15]. For their analysis and theoretical description, a concept of plastic strain-induced PTs was introduced in[16,17]. Traditional temperature-, pressure-, and stress-induced PTs predominantly initiate at pre-existing defects (various dislocation configurations, stacking faults, disclinations, etc.), which produce a concentration of stress and nucleation sites. For stress-induced PTs, applied stresses are below the yield strength. Because of the limited number of pre-existing defects, one must increase pressure/stress and/or change temperature to activate nucleation at defects with weaker stress concentrations. PTs caused by plastic flow are coined as “plastic strain-induced PTs.” They occur by nucleating at new defects permanently generated during plastic flow. The only defect that can reduce the PT pressure by one to two orders of magnitude is the dislocation pile-up. It possesses the largest concentration of all components of the stress tensor at its tip, proportional to the number of dislocations in a pile-up, which can be 10 or even 100. Strain-induced PTs represent a separate class of PTs that require a completely different thermodynamic and kinetic description, as well as experimental characterization. The dislocation pileup-based nucleation mechanism has been confirmed by analytical[16,17], phase-field[18,19], and atomistic[20,21] approaches, as well as experimentally[3]; see reviews[13-15].

Compression in a diamond anvil cell (DAC) and especially torsion in an RDAC represent methods for in situ studies of strain-induced PTs[2,3,5,13-15,22], in contrast to ex situ studies using high-pressure torsion (HPT) with metallic/ceramic anvils[4,6,7]. It is generally accepted that for strain-induced PTs, the volume fraction $c$ of the high-pressure phase (HPP) is a growing function of the plastic strain; when straining stops, PTs also stop[2,4,13-17]. Therefore, time is not an essential parameter; instead, plastic strain plays the role of a time-like parameter. Based on this assumption, the first general kinetic equation for strain-induced PTs was derived in[16,17] using the accumulative plastic strain $q$ as a time-like parameter. It has been experimentally confirmed in[22] and further advanced via a coupled experimental-computational method in[23],[24] for irreversible $\alpha$-$\omega$ PT in Zr only. Additionally, rules governing the evolution of microstructure (crystallite size and dislocation density) during SPD and PTs, as well as the mutual effects of the microstructure and PTs, are found in situ for Zr[25,26] and Si[3]. All these RDAC studies have been performed under slow quasi-static loading; in HPT, the maximum achieved strain rate was 20/s[27], but without in situ capabilities.

However, practical and fundamental needs require studying processes in RDAC at high strain rates, up to $10^4$ $s^{-1}$. Many applications, such as military (armor penetration, shock propagation, and mitigation[10]), dynamic material synthesis, ball milling, high-velocity friction and wear, deep-focus earthquake initiation[11,12], meteorite impact, etc., require understanding material behavior under high-strain-rate SPD and PT. Fundamental questions include whether strain rate and time influence strain-induced PT and microstructure evolution and, if so, how. What new rules govern these phenomena, and are new theories required? To address these problems, we developed a dRDAC with rotation speeds up to 4,000 RPM and achieved strain rates up to $\sim 2{,}299/s$ pressures up to 27.6 GPa, and SPD. None of the other dynamic methods, such as drop-weight impact, dynamic DAC, Kolsky/Hopkinson bars, plate, pressure-shear plate, laser impacts, shock waves, Z-machine, etc., can provide SPD under high pressure within this parameter range, along with comprehensive in situ probing of the processes occurring. In addition, since the temperature does not change during deformation due to the high thermal conductivity of diamonds and the small sample thickness, the strain-rate effects are decoupled from temperature, which is not possible with other techniques.

The Fe-7%Mn alloy is chosen as the first material for testing due to its importance for shock mitigation strategies[10] and its unique transformational and deformational properties[28,29]. It undergoes reversible BCC (SG: $Im\bar{3}m$) to HCP (SG: $P6_3/mmc$) PT, and plastic straining strongly affects both PTs. Therefore, a unique scenario occurs in which both direct and reverse PTs occur simultaneously until steady $c_s$ is reached, which differs with pressure. The mechanism and kinetics of such a PT are expected to be markedly different from those of irreversible strain-induced PTs[16,17] but have never been studied quantitatively in experiments for any material.

## Experimental Results

Under plastic compression, the BCC→HCP PT is dramatically reduced to ~3.2 GPa (Fig. 2a and Fig. S1a) from 11.4 GPa under hydrostatic compression[30]. For comparison, for Fe, the PT initiation pressure is 14.9-15.5 GPa under hydrostatic loading and 10.9 GPa during torsion in RDAC[31].

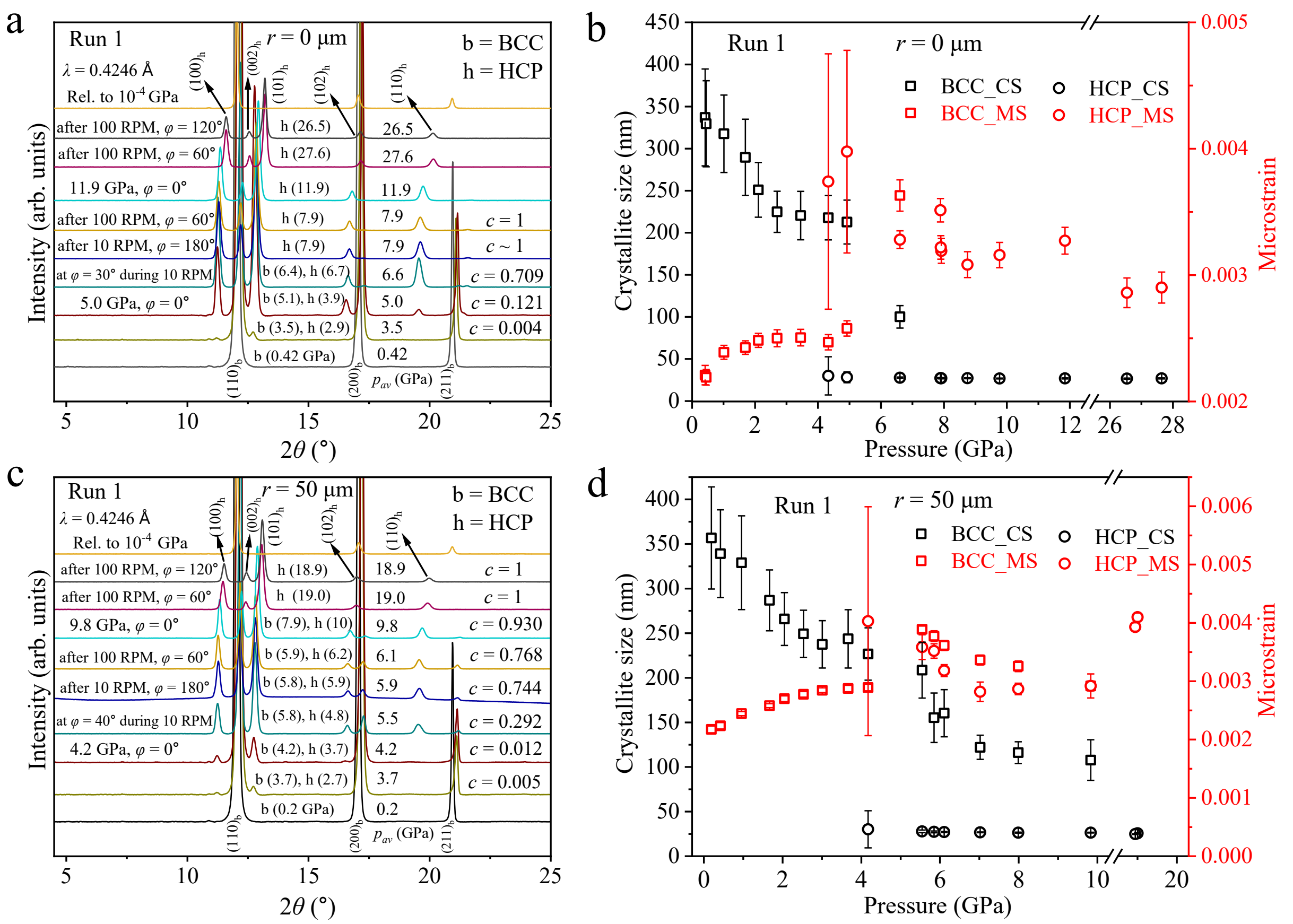

**Fig. 1. High-pressure XRD patterns and microstructure evolution of the Fe-7%Mn alloy collected after compression and torsion in dRDAC at different rotation rates in run 1.** The in situ data during shear are provided only for $\varphi = 40^\circ$, 10 RPM; **(a)** and **(c)** are XRD patterns at $r = 0$ and $r = 50$ μm, and **(b)** and **(d)** show the microstructure evolution at $r = 0$ and $r = 50$ μm. The compression torsion program, including rotation rate in RPM and rotation angle ($\varphi$) for the

corresponding rotation rates, is shown on the left side. In the next column, the pressures in the BCC (b) and HCP (h) phases are shown in parentheses. The average pressure ($p_{av}$) is shown in the middle-right column. The volume fraction *c* of the HCP phase is presented in the rightmost column. Pressure values in the figure are rounded to one decimal place for clarity; the corresponding original values, with their errors to two decimal places, are provided in the supplementary tables.

Figs. 1 and 2 exhibit the evolution of HPP, crystallite size, and microstrain in phases under compression and shear at the sample center, *r* = 0 µm, and at the radius *r* = 50 µm for runs 1 and 2. In run 1, the sample was initially compressed to 5.0 GPa, resulting in *c* = 0.121(4) at *r* = 0 µm (Fig. 1a). The initiation of the HCP phase at *r* = 0 µm is noticed by the appearance of two shoulder peaks, $(100)_h$ and $(101)_h$, on either side of the $(110)_b$ peak of the BCC phase at 3.5 GPa, which is 3.7 GPa at $r$ = 50 μm (Figs. 1a and c). A 180° shear was then applied at 10 RPM under a pressure of 5.0 GPa at $r$ = 0 μm. For the 10 RPM experiment, we collected time-resolved XRD data along the sample diameter. During torsion at 10 RPM by 40° (during which 20 data points were collected in the range -150 μm ≤ $r$ ≤ 50 μm, with a 2° rotation increment between points), the average pressure over both phases $p_{av}$, at $r$ = 50 μm, increases from 4.2 to 5.5 GPa, accompanied by an increase in $c$ from 0.012(4) to 0.292(7) (Fig. 1c). But at $r$ = 0 μm, $c$ drastically increases from 0.121(4) to 0.709(16) for rotation angle $\varphi$ = 30° and $p_{av}$ increases from 5.0 to 6.6 GPa. After $\varphi$ = 180°, at $r$ = 50 μm, $p_{av}$ increases to 5.9 GPa and $c$ = 0.744(23), while at the center, $p_{av}$ increases to 7.9 GPa, and PT is practically complete (Supplementary Tables 1-2).

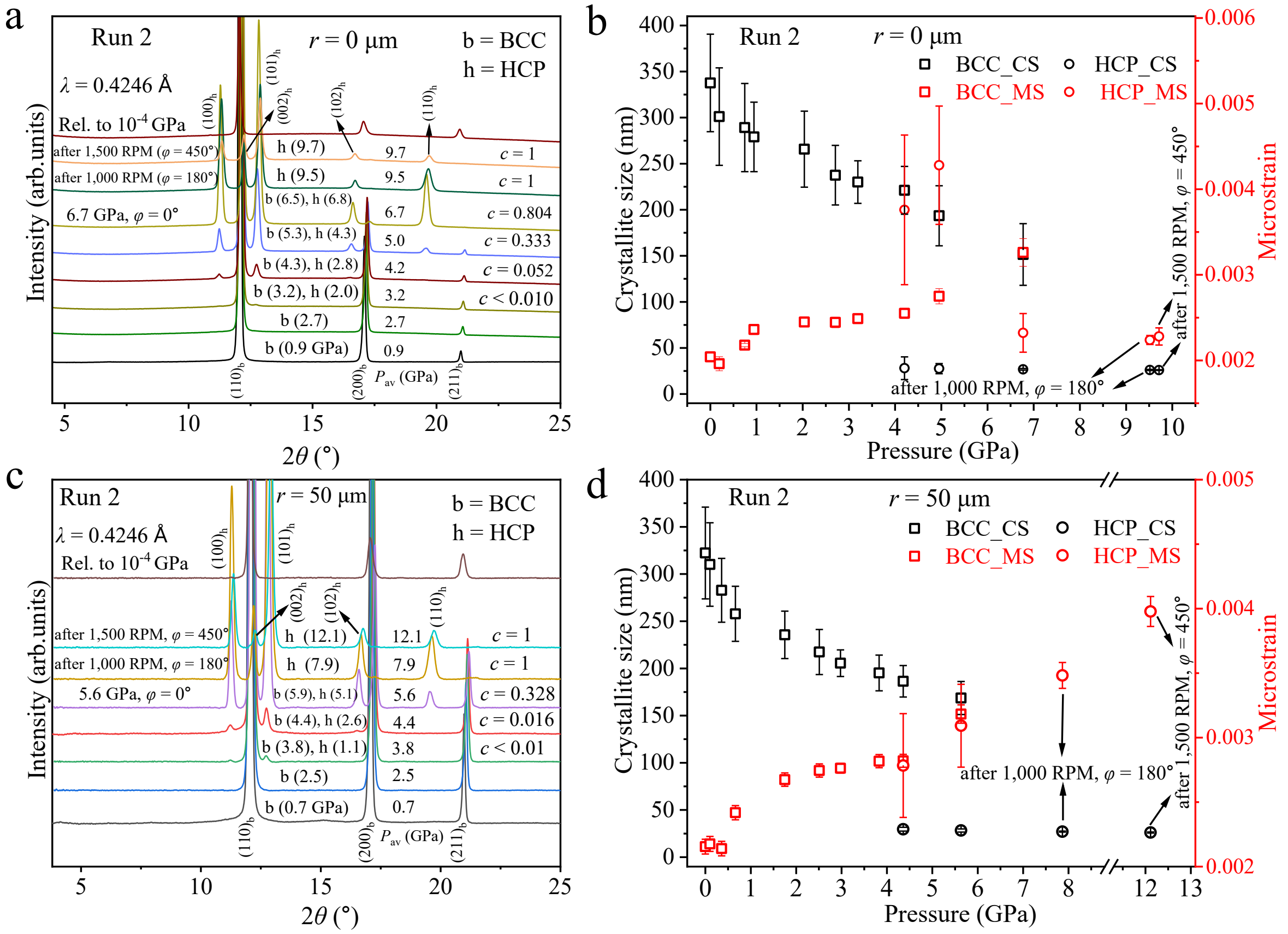

**Fig. 2. High-pressure XRD patterns and microstructure evolution of the Fe-7%Mn alloy collected during and after compression and torsion in dRDAC at different rotation rates for run 2. (a)** and **(c)** HPXRD at *r* = 0 and *r* = 50 µm for run 2, and **(b)** and **(d)** the crystallite size and microstrain evolution at *r* = 0 and *r* = 50 µm.

For $r$ = 50 μm, torsion by 180° at 10 RPM and subsequent torsion by 60° at 100 RPM, the change in pressure in the BCC phase is negligible at $r$ = 50 μm, whereas pressure in the HCP phase increases by 1.35 GPa, becoming practically equal to the pressure in the parent phase (Supplementary Table 2). However, both phases show a significant pressure increase at $r$ = 0 μm (Fig. 1a, c). Therefore, large plastic shear eliminates the internal pressure difference between phases that arises from transformation strain. The initiation of BCC-HCP PT was observed at 3.2 GPa, $r$ = 40 μm (3.5 GPa at $r$ = 0 μm) (Fig. 2 and Fig. S1a). The sample was further compressed to 11. 9 GPa at $r$ = 0 μm. As a result, the average pressure increased to 9.8 GPa at $r$ = 50 μm. A pressure increase of about ~2.1 GPa is observed in the HCP phase relative to the BCC phase (Fig. 1c). Also, $c$ increases to 0.930(31), as evidenced by the reduction in the intensity of the $(200)_b$ peak of the BCC phase. At 11.9 GPa, at $r$ = 0 μm, the sample was again sheared by 60° at 100 RPM, resulting in a significant pressure increase from 11.9 to 27.6 GPa (Fig. 1a). At $r$ = 50 μm, the BCC-HCP PT is completed, and the pressure is increased from 9.8 to 19.0 GPa (Fig. 1c). Upon applying an additional 120° shear at 100 RPM, the pressure at $r$ = 0 μm is dropped by 1.1 GPa. In contrast, the pressure remains constant at $r$ = 50 μm. Time-resolved XRD measurements were conducted at $r$ = 0 μm for all 100 RPM shear experiments at 7.9, 11.9, and 27.6 GPa at $r$ = 0 μm. A reverse HCP-to-BCC PT is initiated at 4.5 GPa, as evidenced by the emergence of the $(200)_b$ peak (Fig. S1b), and is completed at 1.4 GPa (Fig. S1c).

In Run 2, the sample was initially compressed to 6.7 GPa in several steps, exhibiting $c$ = 0.804(38) at $r$ = 0 μm (Fig. 2a). At $r$ = 50 μm, the pressure and $c$ for the same compression stage are 5.6 GPa and 0.328(9), respectively (Fig. 2c and Supplementary Table 4). The initiation of BCC-HCP PT is seen at 3.2 GPa at $r = 0$ µm and 3.8 GPa at $r = 50$ µm (Fig. 2a, c). At 4.2 GPa, the pressure in the BCC phase is 1.5 GPa higher than that in the HCP phase at $r = 0$ µm, whereas it is 2.7 GPa ($p_{av} = 3.8$ GPa) higher at $r = 50$ µm (Fig. 2a, c and Supplementary Tables 3, 4). After applying a 180° shear at 1,000 RPM and 6.7 GPa at $r = 0$ µm, the BCC phase completely transforms into the HCP phase at $r = 0$ and $r = 50$ µm, with the pressure increasing to 9.5 GPa and 7.9 GPa at these points, respectively.

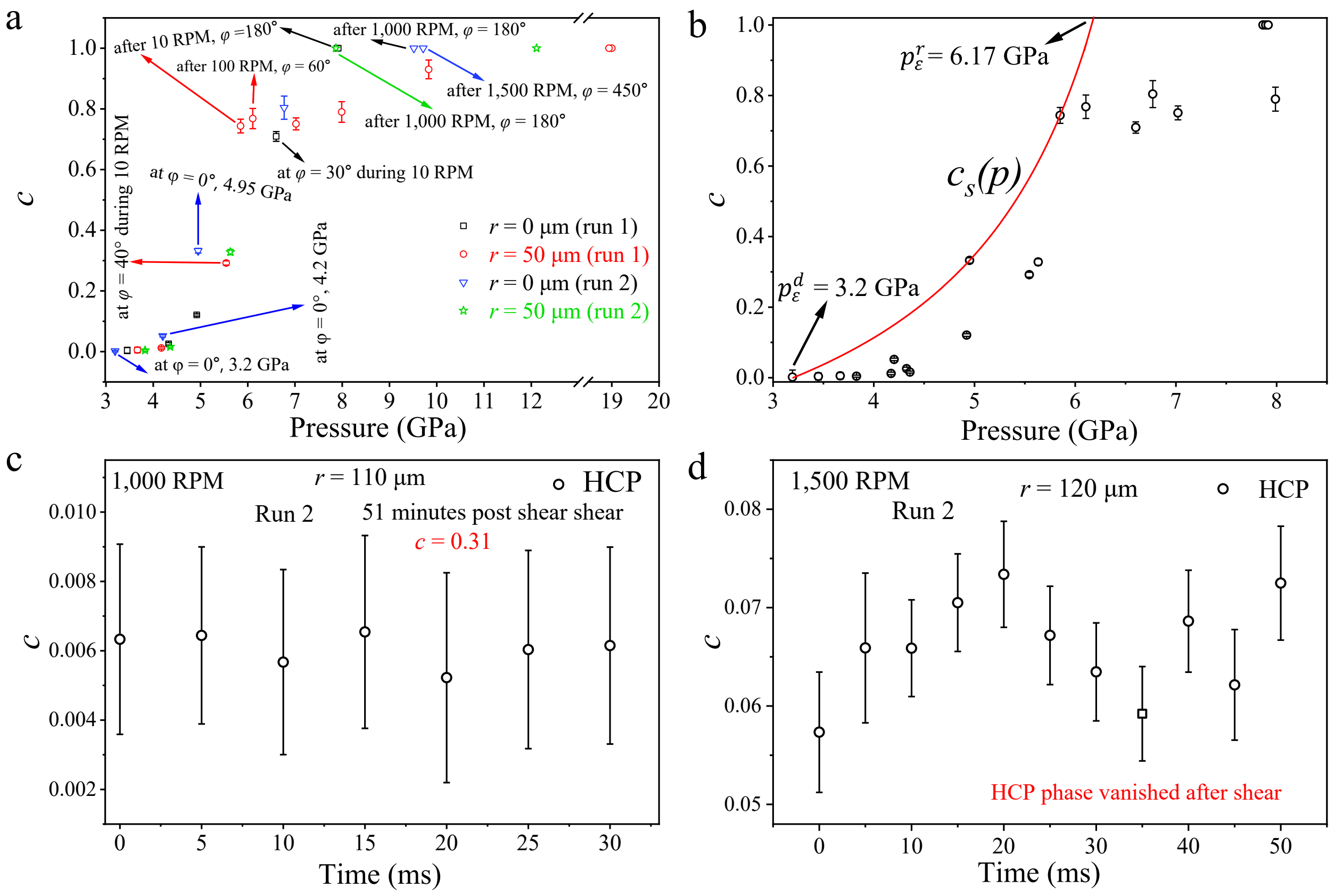

**Fig. 3. Evolution of the volume fraction (*c*) of phases for different compression-torsion loadings before, during, and after shear at different radii, *r*. (a)** At $r = 0$ µm and $r = 50$ µm for runs 1 and 2 after compression and torsion with different rotation rates as a function of pressure. **(b)** The same data for a smaller pressure range and theoretical curve (red) for the steady volume fraction of the HCP phase $c_s(p)$ from Eq. (2) fitted to the three points with maximum $c$ for each pressure. **(c)** Time-resolved evolution of $c$ with a 5 ms step size at $r = 110$ µm, 1,000 RPM, and at 6.7 GPa at $r = 0$ µm. **(d)** Time-resolved evolution of $c$ with a 5 ms step size at $r = 120$ µm, 1,500 RPM, and at 9.5 GPa at $r = 0$ µm.

When the sample was subsequently sheared by 450° at 1,500 RPM, only a slight pressure increase of 0.2 GPa was seen at $r = 0$ μm, whereas a pronounced pressure increase of 4.2 GPa was observed at $r = 50$ μm. Time-resolved XRD data were collected at $r = 110$ and $r = 120$ μm during shearing at 1,000 and 1,500 RPM, respectively. At 1,500 RPM, the pressure in the HCP phase at $r = 50$ μm is 2.4 GPa higher than at $r = 0$ μm. The volume, volume fractions, crystallite size, and microstrain obtained from the real-time XRD measurements are shown in Supplementary Tables 3-4.

The weak reflection near ~11° in Run 1 could be attributed to an unknown impurity introduced into the sample during sample loading. We do not see this reflection in Run 2. We also have not observed this peak in our hydrostatic XRD studies on Fe-7%Mn with He as PTM[30]. Our XRD analysis indicates that the additional reflection does not correspond to any known Fe oxide phases. Furthermore, we performed an additional XRD measurement on another well-cleaned Fe-7%Mn sample at an X-ray energy of 0.3445 Å at 16-ID-B, and we did not observe any reflection near 11° (Fig. S6). Moreover, we observed a PT initiation pressure of 3.2 GPa for BCC-HCP PT in both Run 1 and Run 2. Hence, we conclude that the observed PT is an inherent property of the sample and that impurities do not influence the phase transformation.

Figs. 1b, d, and 2b, d show the evolution of crystallite size (CS) and microstrain (MS) for the BCC and HCP phases. CS and MS evolve in opposite directions until the onset of the PT. Since the XRD peaks of the HCP phase are too weak at the onset of the PT, the CS and MS values at the initiation point are not reported. The maximum change in the crystallite size of the HCP phase is approximately 5.4 nm in Run 1 and 3.5 nm in Run 2, whereas it is 249 and 128 nm for the BCC phase in Runs 1 and 2, respectively (Supplementary Tables 8-11).

Figures 3a and 3b show most of the representative experimental data from both runs, plotted as $c$ versus $p_{av}$ over different pressure ranges. These data are used below to connect with the theory and determine the steady-state volume fraction $c_s(p)$ (red line in Fig. 3b) and material parameters in the kinetic equation for strain-induced PT. For the time-resolved data in Fig. 3c, during torsion at 1,000 RPM and $r = 110$ μm, the pressure is 3.65 GPa in the BCC phase and 2.93GPa in the HCP phase before torsion (Supplementary Table 6) and remains essentially unchanged during torsion. $c$ varies within a narrow range of 0.0052(30)-0.007(3), i.e., PT essentially does not occur during torsion within the temporal resolution of the experiment (Fig. S3a-b). Surprisingly, $c$ increases to 0.31(2) at 51 minutes after torsion stops at 4.05 GPa (BCC) and 4.53 GPa (HCP) (Fig. S3a-b, Supplementary Table 6). During the subsequent torsion at 1,500 RPM and $r$ = 120 μm, the pressure is 3.64 GPa in the BCC phase and 3.65 GPa in the HCP phase before torsion and remains basically unchanged during torsion (Supplementary Table 7). The $c$ varies in a range of 0.066(7)-0.073(5) during the torsion (Fig. 3d). After torsion stops, $p_{av}$ = 3.95 GPa, and the HCP phase disappears after 7 minutes. Figs. S2 and 3a-d and Supplementary Tables 6-7 show the evolution of the HCP phase and the volumes of the BCC and HCP phases during torsion at 100, 1,000, and 1,500 RPM.

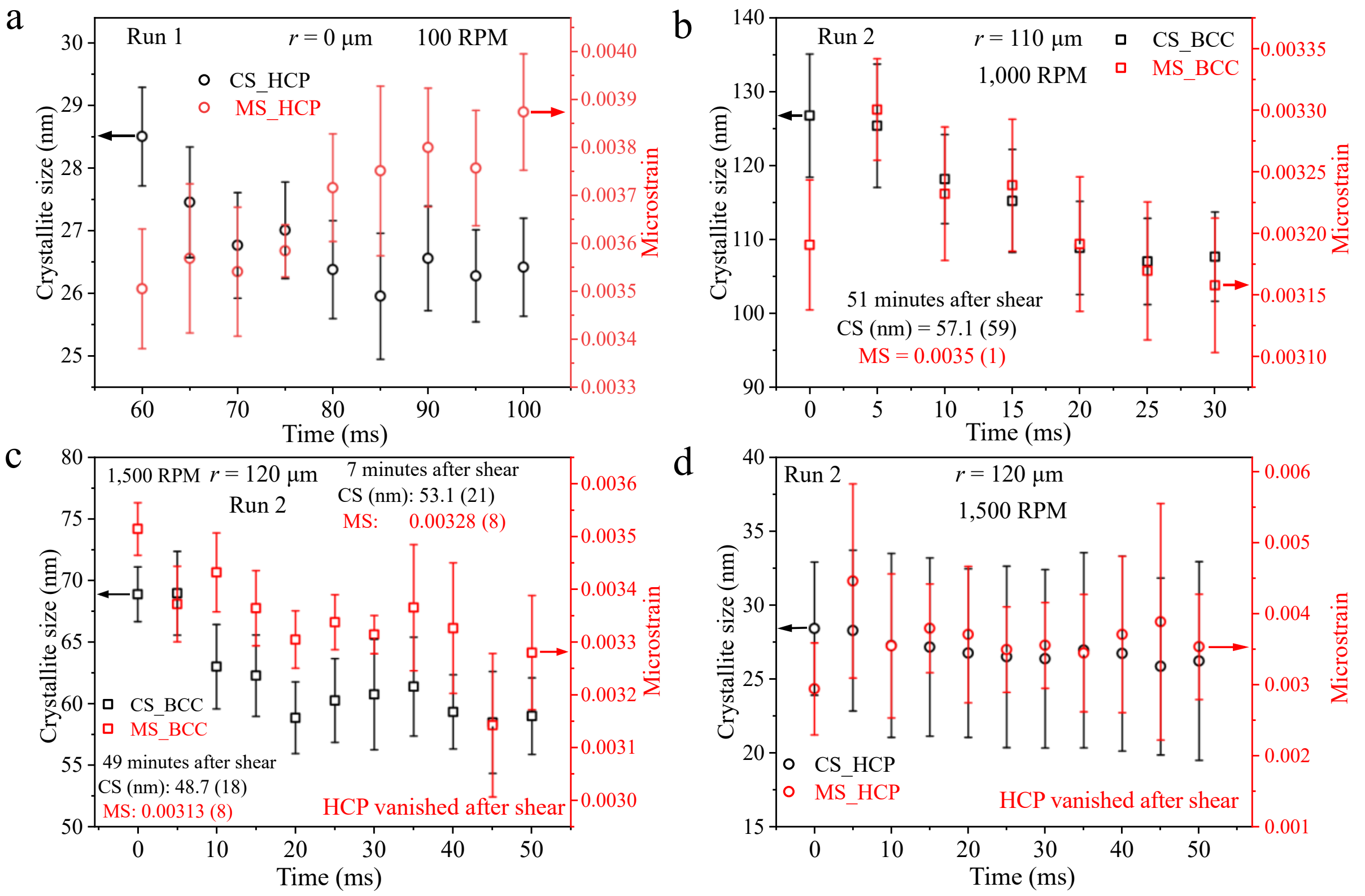

**Fig. 4. Time-resolved XRD study of crystallite size and microstrain evolution in BCC and HCP phases during the same compression-torsion loadings as in Figs. 3c and d, and for 100 RPM at *r* = 0.** Microstructure evolution of **(a)** in the HCP phase at $r$ = 0 μm for 100 RPM at 7.9 GPa at $r$ = 0 μm, $\varphi$ = 60°; **(b)** in the BCC phase at $r$ = 110 μm for 1,000 RPM; **(c)** in the BCC phase at $r$ = 120 μm for 1,500 RPM, and **(d)** in the HCP phase at $r$ = 120 μm for 1,500 RPM. Each time-resolved data was collected with a step size of 5 ms.

During compression, the crystallite size in the BCC phase reduces from 338(53) to 151.5(33) nm at 6.7 GPa at $r$ = 0 μm and to 169(18) nm at 5.6 GPa at $r$ = 50 μm in run 2 (Figs. 2b and d). The crystallite size in the HCP phase is ~28(7) nm from the onset of nucleation and remains unchanged during subsequent static compression for both sample radii, during torsion at 1,000 and 1,500 RPM, and after each torsion (Figs. 2b, d, and 4d). The microstrain in the HCP phase remains essentially constant during these processes, equal to ~0.0035(1). Since the dislocation density in the Rietveld refinement procedure[1,2] is determined in terms of crystallite size and microstrain, it also remains constant in the HCP phase. Using a simplified equation for the dislocation density $\rho_d = \frac{1}{d^2}$, we obtain $\rho_d = 1.3(6) \times 10^{15}$/m$^2$.

The time-resolved evolution of the crystallite size and microstrain in the BCC phase during the same loadings as in Fig. 4d is presented in Figs. 4b and c. For 1,000 RPM, the microstrain increases from 0.00319(5) to 0.00330(4) during acceleration, then reduces from 0.00317(6) to 0.00316(5) (practically no change) during deceleration to zero rotation speed. The crystallite size reduces from 108(6) nm during deceleration to zero rotation speed to 57.1(59) nm, and the microstrain increases from 0.00316(5) to 0.0035(1) after 51 minutes of torsion stopping. For 1,500 RPM (Fig. 4c), the crystallite size in the BCC phase decreases from 69(3) nm to a steady ~59(3) nm. The microstrain decreases from 0.00351(5) to 0.00337(7) during acceleration, then increases from 0.00314(20) to 0.00328(11) during deceleration to zero rotation speed. During torsion, the microstrain reduces from 0.00337(7) to 0.00314(8). After torsion stops, they change to 53.1(21) nm and 0.00328(8) after 7 minutes and to 48.7(1.8) nm and 0.00313(8) after 49 minutes. For 100 RPM in run 1 (Fig. 4a), during torsion, the crystallite size decreases from 28.5(79) to 26.4(78) nm, whereas the microstrain increases from 0.00351(13) to 0.00387(12). After torsion, they change to 27.1(48) nm and 0.00321(9). After the pressure release, the crystallite size and microstrain of the BCC phase are 61.7(25) nm and 0.00300(8), respectively.

Results of our finite element method (FEM) simulations in Figs. S11 and S12 conclude that the maximum temperature increment in the sample at 1,500 RPM does not exceed 1 K, i.e., is negligible. Since the maximum temperature increment is proportional to $\tau_y \omega$, an increase in this product, even by a factor of 10 or more, does not affect PTs and microstructure evolution (Supplementary Discussion Section VI).

The in situ optical imaging of the Fe-7%Mn was performed at 1,000 RPM, $p$ = 4.3 GPa, at $r$ = 0 μm, and $\varphi$ = 120° with a Photron FASTCAM SA3 camera. Digital image correlation using the open-source 2D subset-based software package Ncorr shows that the imposed anvil rotation angle (120°) closely matches the sample rotation angle (120.7°), with a deviation of 0.6%, suggesting no slippage (Supplementary Discussion Section IV and Supplementary Movies 1 and 2).

The sample thickness along various compression and shear *p-q* paths was calculated, which allowed us to estimate increments of plastic shear strain $\Delta\gamma_{avg}$, logarithmic axial strain $\Delta\varepsilon_a$, maximum shear rate $\dot{\gamma}_{max}$, and averaged axial strain rate $\dot{\varepsilon}_a$ (Supplementary Tables 12 and 13 and Supplementary Discussion Sections II and V). Interestingly, a large axial strain ($\varepsilon_a = 2.97$) is obtained at a low pressure of 9.52 GPa after $\varphi$ = 180° at 1,000 RPM and $\varphi$ = 450° at 1,500 RPM, compared to a strain of $\varepsilon_a = 2.66$ at a high pressure of 26.54 GPa after multiple compressions and torsion by $\varphi$ = 180° at 10 RPM and total $\varphi$ = 240° at 100 RPM (Supplementary Tables 12 and 13). Shear strain during torsion is much larger; e.g., $\Delta\gamma_{avg}$= 71.94 for $\varphi$ = 450° at 1,500 RPM. The maximum strain rate is 640/s during rotation at 1,000 RPM and 2,299/s at 1,500 RPM.

**Discussion**

**Quasi-static kinetics of strain-induced PTs**

According to the theory in[16,17], when both direct and reverse PTs occur simultaneously, the kinetic equation for the strain-induced PT is

$$\frac{dc}{dq} = \frac{k_d(1-c)}{c\left(\frac{\sigma_{y1}}{\sigma_{y2}}\right)+(1-c)}\left(\frac{p-p_\varepsilon^d}{p_h^d-p_\varepsilon^d}\right) H(p-p_\epsilon^d) - \frac{k_r c}{c+(1-c)\left(\frac{\sigma_{y2}}{\sigma_{y1}}\right)}\left(\frac{p_\varepsilon^r-p}{p_\varepsilon^r-p_h^r}\right) H\,(p_\varepsilon^r-p). \qquad (1)$$

Here, $p_\varepsilon^d$ and $p_\varepsilon^r$ are the minimum pressure for direct and maximum pressure for the reverse strain-induced PTs, respectively; $p_h^d$ and $p_h^r$ are the minimum pressure for direct and maximum pressure for the reverse pressure-induced PTs, respectively; the Heaviside unit step function *H* (*H(x) = 1* for *x* ≥ *0; H(x) = 0* for *x* < *0*) ensures that the term for the direct PT contributes at $p > p_\varepsilon^d$ only, whereas the term for the reverse PT is nonzero for $p < p_\varepsilon^r$; $k_d$ and $k_r$ are the kinetic coefficients for direct and reverse PTs, respectively; $q$ is the accumulated plastic strain, and $\sigma_{y2}$ and $\sigma_{y1}$ are the yield strengths of the high- and low-pressure phases, respectively. Eq. (1) is derived utilizing results from the nanoscale mechanism of the strain-induced nucleation at the tip of the dislocation pile-up[16,17]. Since edge dislocation pile-ups generate both compressive and tensile stresses of the same magnitude in different regions, they promote simultaneous direct and reverse PTs in different regions. Therefore, Eq. (1) has terms for direct and reverse PTs. Eq. (1) is independent of time; plastic strain $q$ is the time-like parameter.

In the pressure range $p_\varepsilon^d \le p \le p_\varepsilon^r$ (i.e., in the very unusual case when pressure $p_\varepsilon^d$ for the direct PT to an HPP is lower than the pressure $p_\varepsilon^r$ for the reverse PT), Eq. (1) has a stationary solution

$$c_s = \frac{1}{1+M\frac{(1-\tilde{p})}{\tilde{p}}}; \qquad \tilde{p} = \frac{p-p_\varepsilon^d}{p_\varepsilon^r-p_\varepsilon^d}; \qquad M = \left(\frac{\sigma_{y1}}{\sigma_{y2}}\right)\left(\frac{k_r}{k_d}\right)\frac{\left(p_h^d-p_\varepsilon^d\right)}{\left(p_h^d-p_\varepsilon^d\right)}, \tag{2}$$

which can be reached for large enough plastic strain; it is an increasing function of pressure; $c_s(p_\varepsilon^d) = 0$ and $c_s(p_\varepsilon^r) = 1$. For $p > p_\varepsilon^r$ and $p < p_\varepsilon^d$, complete direct and reverse PTs can be reached in a steady state. While Eq. (1) for the absent reverse PT was calibrated for $\alpha \rightarrow \omega$ PT in Zr[22-24], the rare case with simultaneous direct and reverse PTs was not considered for any material.

In Figs. 3a and b, all experimental points for different plastic strains are placed in the *p-c* plane. Points with maximum *c* for each *p* can be considered a lower bound for the steady $c_s(p)$; see the red line in Fig. 3b. One point on this line, $p_\varepsilon^d = 3.20$ GPa, with $c_s(p_\varepsilon^d) = 0$, is determined directly from experiment. Eq. (2) for $c_s(p)$ has two more material parameters, $p_\varepsilon^r$ and *M*. They are determined from the condition that the plot $c_s(p)$ passes through two other points with maximum *c* for all pressures (Fig. 3b). This leads to $p_\varepsilon^r = 6.17$ GPa and *M* = 2.90. Note that these parameters were not previously known for any material.

Analysis of Eq. (2) shows that for *M* = 1, $c_s(M = 1) = \tilde{p}$, which indicates equal promotion of direct and reverse PTs. Since $c_s$ is a decreasing function of *M*, obtained *M* = 2.90 means that the reverse PT is promoted more than the direct one. Given the known M, let us evaluate the material parameters that affect it. Taking iron $\sigma_{y1} = 1.1$ GPa and $\sigma_{y2} = 3$ GPa[32] and assuming that the modest effect of Mn on the strength of Fe[10] does not change the ratio $\sigma_{y1}/\sigma_{y2}$ for iron, we obtain $\sigma_{y1}/\sigma_{y2} = 0.367 < 1$. Such a value smaller than 1 promotes direct PT by localizing the applied plastic strain in a weaker, lower-pressure phase.

For $p_\varepsilon^d = 3.20$ GPa, $p_\varepsilon^r = 6.17$ GPa, $p_h^d = 11.4$ GPa, and $p_h^r = 4.5$ GPa, we have $\frac{p_h^d-p_\varepsilon^d}{p_\varepsilon^r-p_h^r} = 4.91 > 1$, i.e., this parameter promotes reverse PT. It looks counterintuitive that a larger effect of the plastic straining on PT pressures $p_h^d - p_\varepsilon^d$ for direct PT than on $p_\varepsilon^r - p_h^r$ for reverse PT promotes reverse PT kinetics. However, this follows from Eq. (1), where these differences appear in the denominators of the corresponding PT terms. With these values and $M = 2.90$, one obtains $\frac{k_r}{k_d} = 1.61$, which makes an additional contribution to the promotion of the reverse PT compared to the direct PT. Thus, the essential volume fraction of the HCP phase can be obtained at a pressure essentially higher than $p_{\varepsilon}^{d}$ only. Thus, all parameters in the kinetic Eq. (1), except one (e.g., $k_d$ or $k_r$), are estimated; a specially designed experiment is required to complete the kinetic model. Since the fastest experiment for which data for $c_s$ were used was at 10 RPM and small $r$ = 50 μm, the obtained kinetic parameters characterize quasi-static loading.

**Need for time-dependent kinetics of strain-induced PT**

The main surprise is that PT practically does not occur during fast torsion at 1,000 and 1,500 RPM, but $c$ increases by 0.31(2) after torsion at 1,000 RPM in 51 minutes and decreases by 0.07 to zero after torsion at 1,500 RPM in 7 minutes. These results challenge the traditional wisdom that the kinetics of strain-induced PTs are independent of time and depend on plastic strain only as a time-like parameter. Theoretically, this statement was justified in[16,17] as follows. Nucleation of the HPP at the tip of the dislocation pile-up is barrierless and occurs on a picosecond to nanosecond timescale. Growth away from the tip is very limited because stress drops sharply with distance from the tip. Usually, it occurs until a steady state is reached, which typically happens within a time frame much shorter than the measurement time. Consequently, only steady states are detected, which depend solely on plastic strain and are independent of time. Here, due to much faster deformation, smaller time resolution, and low pressure, time-dependent growth at constant plastic strain is revealed. Thus, HCP nuclei are generated during straining at high strain rates, but their growth/disappearance until steady state occurs on the few-minute time scale. Since growth occurs at constant plastic strain, it represents stress-induced PT, i.e., a different mechanism. Generally, both strain- and stress-induced PTs occur simultaneously in different parts of the representative volume, necessitating a fundamentally new theory for combined strain- and stress-induced PTs that considers strain rate and time effects, as well as evolving microstructure. The results obtained represent the first experimental findings in this direction.

Formal nanoscale[18,19] and microscale[33,34] phase-field approaches consider combined strain- and stress-induced PTs by studying coupled stress-induced dislocation evolution and PTs. However, they are not applied to or calibrated with experimental data, and macroscale kinetics are not developed. Still, nucleation and growth stages are clearly seen in[18,19], where growth in nanograined materials is promoted by a small distance between different nuclei, their interaction, and coalescence, which the analytical approach[16,17] neglects. Since pressure during time-resolved torsion at 1,000 RPM (3.65 GPa in BCC, 2.93 GPa in HCP, and $p_{av}$ = 3.65 GPa) before torsion is very close to $p_\varepsilon^d$, growth is very slow. After torsion stops, pressure is higher (4.05 GPa in BCC and 4.53 GPa in HCP (Supplementary Table 6)) but still close to $p_\varepsilon^d$, and the growth takes a few minutes. For torsion at 1,500 RPM, pressure is 3.64 GPa in BCC and 3.65 GPa in HCP (Supplementary Table 7) before torsion, which is even closer to $p_\varepsilon^d$. Actually, it is much below $p_h^r = 4.5$ GPa, and after plastic straining stops and shear stresses are partially relaxed, pressure-induced reverse PT may occur within a few minutes. After the reverse PT completes, $p_{av}$ = 3.95 GPa, but without plastic straining, this does not cause direct PT. Also, variations in the microstructure and microstrain (Fig. 4 b- d) and relaxation of internal stresses, which differ in different regions, may contribute to HCP phase growth after rotation at 1,000 RPM and disappearance after twisting at 1,500 RPM.

Also, as follows from nanoscale phase-field simulations for a bicrystal[18,19] and microscale simulations for a polycrystal[33,34], the steady HPP state obeys the PT condition

$$< \sigma_{ij} > \varepsilon_t^{ij} = \Delta G^\theta, \quad (3)$$

where $< \sigma_{ij} >$ and $\varepsilon_t^{ij}$ are the averaged polycrystal stress and transformation strain tensors, and $\Delta G^\theta$ is the difference in the thermal part of the free energy between phases. This is the same condition as for stress-induced PTs. Since averaging filters out internal stresses, the steady phase state is governed by the applied stress tensor or, in a simplified version, by the applied pressure. The strain-induced contribution produces nucleation at low pressures, where stress-induced nucleation is impossible. The plot in Fig. 3b is the first experimental confirmation of the prediction that pressure determines the steady volume fraction for combined strain- and stress-induced PT.

As follows from the theory and simulations on nucleation at the tip of a dislocation pile-up, stresses/pressure in the small nuclei of a high-pressure phase should be significantly larger than applied pressure (despite the volume reduction during PT), which was confirmed experimentally for PTs Si I→Si II and Si I→Si III[3]. This is not the case for Fe-7%Mn for small *c* (Figs. 1a, c, and 2a, c), similar to that for *α*-*ω* PT in Zr[22]. Because HPP has 12 martensitic variants, variant-variant transformations, which often represent twinning, should significantly relax deviatoric stresses during the PT. The maximum driving force for the PT is achieved at zero deviatoric stress after completing the PT in nuclei[35], which should also reduce pressure in nuclei. For large *c*, pressure in the HCP phase is getting larger than in the FCC phase, but this is due to the higher yield strength of the HCP phase. At the same time, as mentioned above, the pressure in the HCP phase is slightly larger than in the BCC phase during torsion at 1,000 and 1,500 RPM, probably due to the much shorter time for stress relaxation. While the same difference remains after torsion at 1,000 RPM, it may be related to an increase in $c$ by 0.31(2). Additionally, we obtain an interesting result: after 1,500 RPM, the pressure in the complete HCP phase at $r$ = 50 μm is 2.4 GPa higher than at $r$ = 0 μm (Figs. 2a and c). This contradicts the analytical solution in[16,17], which was a basis for the developing intuition for this process. However, more precise finite element method solutions[36,37] show a similar trend, which has yet to be experimentally confirmed.

**Microstructure evolution**

The results obtained in Fig. 4 for the microstructure evolution in the HCP phase led to the new and very important rule:

Crystallite size of ~28(6) nm, microstrain ~0.0035(8), and dislocation density $\rho_d = 1.3(6) \times 10^{15}/\text{m}^2$ in the HCP phase remain steady during static compression and dynamic torsion, during and after the PT, and after torsion, and are independent of pressure, plastic strain tensor, its path, strain rates, and the volume fraction of the HCP phase.

Indeed, during compression and torsion, the stress-strain state of the sample is highly heterogeneous, and each material point undergoes completely different pressure, plastic strain tensor, and its entire path[23,24,36,37]. Therefore, observing the same microstructure at different points in the sample and at different stages of compression and torsion implies independence from pressure, the plastic strain tensor, and the deformation path, and $c$. The same microstructure under static compression and torsion at different rotation rates implies independence from the plastic strain rate. The maximum averaged strain rate for which data were collected, and for which this rule is valid, is ~2,299 $s^{-1}$ (Supplementary Table 13).

Let us compare our results with those of previous studies. Steady grain size and dislocation density and their independence of radius and initial state during slow HPT with SPD are known for various materials[4,6,38-41]. Pressure independence of the steady grain size and/or dislocation density and/or hardness after HPT is found in[42]; their independence of the strain rate in the range from 0.004 to 20/s was determined in[27]. All these results were obtained for a single-phase state after pressure release and ex situ measurements rather than in situ. During unloading, grain size and dislocation density may change by a factor of 2[43,44]. In more precise in situ experiments here, the crystallite size after the pressure release increases from 27.1(48) to 61.7(25) nm (i.e., by a factor of 2.3), and dislocation density reduces from $1.36(48) \times 10^{15}/\text{m}^2$ to $2.63(61) \times 10^{14}/\text{m}^2$ (i.e., by a factor of 5.2). Also, during torsion alone, the plastic strain path is strongly shear-dominated and essentially unchanged. In situ studies on $\omega$-Zr under static compression in a DAC demonstrated that steady crystallite size and dislocation density are independent of pressure, plastic strain tensor, and strain path[25,26], with strong variation in strain paths[23,24]. During the $\alpha \rightarrow \omega$ PT in Zr, the crystallite size and dislocation density in both phases strongly depend on the volume fraction of phases[26]. Here, constant crystallite size and dislocation density in the HCP phase are obtained in situ in a broad strain rate range, from 0 for static compression to $2{,}299/\text{s}$, a broad range of straining paths during compression and torsion, during and after the PT.

Reduction in the crystallite size of the BCC phase during compression before and during the PT in Figs. 2b and d is not surprising. The results in Figs. 4b and c look logical. Indeed, at 1,000 RPM, crystallite size varies monotonously in the range 127(8)-108(6) nm during torsion and then reduces strongly to 57.1(59) nm at constant load, while at 1,500 RPM, crystallite size almost monotonously reduces from 69(3) nm to a steady ~59(3) nm during torsion and slightly reduces to 49 nm at constant load. The potential reason may be the very different *c* values in these cases.

To summarize, the first experiments in dRDAC on SPD, BCC↔HCP PT, and microstructure evolution at pressures up to 27.6 GPa, rotation rates up to 1,500 RPM, and strain rates up to $2{,}299/\text{s}$ are performed with an Fe-7%Mn alloy as an example. Real-time data at specific radial points, as well as along the sample diameter, are collected with acquisition times of 5 and 10 ms and a beam spot size of 1.3 µm × 1.9 µm. It is found that the initiation pressure for the strain-induced direct PT, $p_\varepsilon^d = 3.20$ GPa, is lower than that for the reverse PT, $p_\varepsilon^r = 6.17$ GPa, which is very counterintuitive when compared with those for the pressure-induced PT, $p_h^d = 11.4$ GPa and $p_h^r = 4.5$ GPa. Such a strong effect of plastic straining leads to the simultaneous occurrence of the direct and reverse PTs in the pressure range $3.20$ GPa $< p <$ $6.17$ GPa. This results in a unique strain-controlled kinetic Eq. (1) with a stationary solution (2), which was not studied for any material. Here, we report material parameters for the kinetics of strain-induced direct-reverse PTs and the stationary volume fraction versus pressure under quasi-static loading.

During torsion at 1,000 and 1,500 RPM at low pressure close to $p_{\varepsilon}^{d}$ the volume fraction of the HCP phase does not change. However, after torsion stops, it surprisingly increases by 31% in 51 minutes after 1,000 RPM, and the HCP phase disappears after 1,500 RPM in 7 minutes. These findings contradict the general wisdom that strain-induced PTs occur only during straining, suggesting that time is not a governing parameter and that kinetics is determined by plastic strain rather than time. Thus, nuclei of the HCP phase are generated during straining at a high strain rate, but growth/disappearance occurs under stress on a timescale of a few minutes. Consequently, a new theory of combined strain- and stress-induced PTs is required. Our nanoscale and microscale phase-field approaches for bicrystals[18,19,33,34] could be a good starting point for such a theory at these scales and govern the development of the macroscale kinetics, which should generalize Eq. (1) significantly. With increasing pressure, the PT time decreases, and we expect to detect the essential evolution of $c$ in real time, similar to the microstructure in Fig. 4c.

The following important rule for microstructure evolution is discovered: crystallite size of ~ 28(6) nm, microstrain ~0.0035(8), and dislocation density ~$1.3(6) \times 10^{15}$/m$^2$ in the HCP phase are steady during static compression and dynamic torsion, during and after the PT, and after torsion. These parameters are independent of pressure, the plastic strain tensor, its path, strain rates, and volume fraction of the HCP phase.

The obtained results open fundamental research on combined strain- and stress-induced PTs and microstructure evolution under dynamic SPD and high pressure in dRDAC for various material systems. This also has significant applied potential for understanding and controlling processes in modern technologies and nature.

## Methods

The Fe-7%Mn alloy with properties described in[10] was prepared in the Army Research Laboratory. A single-phase solid solution of Fe-7%Mn was observed, and the exact composition of the material was found to be Fe-7.04%Mn. The initial dimensions of the sample were 36.8 mm × 25.7 mm × 14.9 mm. We reduced the 14.9 mm sample thickness to approximately 5 mm by cutting it into three pieces using electrical discharge machining (EDM) at the Ames National Laboratory machine shop. We further reduced the sample thickness to 280 μm by cross-rolling on the 100T rolling mill in the same machine shop, producing a nanograined structure. For dynamic shear experiments, we used dRDAC from DAC Tools, LLC, IL, as discussed in Supplementary

Discussion Section I. We used a pair of diamonds with a culet size of 300 μm and roughened surfaces with an asperity of ~400 nm (instead of the traditional 10 nm), prepared for us by lapping with Almax Easy Lab to increase the contact friction stress to the maximum possible one, equal to the yield strength in shear. We punched two samples with a thickness of 280 μm using a 5-mm-diameter punch-and-die set. We pre-indented these samples to thicknesses of 168 μm and 159 μm and used them for the dynamic shear measurements in Runs 1 and 2, respectively. We used no pressure-transmitting media for these experiments. Such a 30-fold reduction in thickness is more than sufficient to reach steady crystallite size, microstrain, and, consequently, dislocation density in metals uniformly across the sample[4,6,15,22-26,45], eliminating these parameters from the list of independent variables and drastically simplifying the study and the search for new rules. For sample 1, the uniform steady values of crystallite size, microstrain, and dislocation density are 353(56) nm, 0.001900(22), and $8.0(25) \times 10^{12}/m^2$, respectively. Similarly, for sample 2, the corresponding values are 338(53) nm, 0.002042(58), and $8.8(27) \times 10^{12}/m^2$, respectively. Therefore, we recommend reaching this state to study PTs and microstructure evolution induced by SPD[15].

We performed high-pressure compression and shear time-resolved X-ray diffraction (XRD) experiments at the 16-ID-B beamline, HPCAT, at the Advanced Photon Source, using X-ray wavelengths of 0.4246 Å. We performed the XRD measurements using a PILATUS3 X 2M CdTe detector, with the X-ray beam focused to a spot size of 1.3 μm × 1.9 μm. The samples were compressed in the dRDAC using a membrane. The time profiles for shear were designed in DT Move software (DAC Tools, LLC), and the dRDAC driving motor (CPM-SCHP-2346P-ELSA) was triggered by the PILATUS3 detector *via* a delay generator (Berkeley Nucleonic model 505) to synchronize XRD collection and shear. The minimum data collection time for time-resolved X-ray diffraction and absorption measurements was 5 ms. The Pilatus 3 2M CdTe detector has a time resolution of 5 ms. When using the detector's full active area, the readout time is 4 ms. With a minimum exposure time of 1 ms, we achieve 5 ms resolution, albeit with a 4 ms gap between exposures. If we reduced the active area, we could improve this to 2 ms, but we would use a smaller detector area and could not collect complete Debye-Scherrer rings at high Q. We collected data along the sample diameter with a 10 μm step size. Thus, at 10 RPM, we performed scans along the sample diameter during the real-time XRD experiments, i.e., along the spiral of the twisted sample. Before and after each torsion, we collected data along the sample diameter. Real-time

XRD measurements at selected data points at the sample center ($r = 0$ μm) and at positions away from the center were collected continuously at anvil speeds of 100, 1,000, and 1,500 RPM. The 2D XRD images were converted into intensity vs. $2\theta$ profiles using Dioptase software[46] and post-processed with Materials Analysis Using Diffraction (MAUD) to calculate the lattice parameters, volume fractions, crystallite size, and microstrain by Rietveld refinement[47].

We analyzed crystallite size (CS) and microstrain (MS) by Rietveld refinement in Materials Analysis Using Diffraction (MAUD) software. We calculated the microstructural parameters of the BCC and HCP phases in Fe-7%Mn samples by slicing the XRD image into 72 sectors with a $5^{\circ}$ step size. We used the $CeO_2$ calibrant to calculate instrument broadening. We fitted the XRD peak broadening by refining CS and MS with an isotropic size-strain model in MAUD. We initially refined the CS and MS parameters independently, then simultaneously (when required) to prevent divergence or over-convergence during refinement. Unreliable CS and MS show abnormal error changes, which we closely monitor at each refinement cycle. The uncertainties in the refined lattice parameters, volume fractions, CS, and MS values were estimated from the variance-covariance matrix generated using a least-squares fitting procedure, with the standard uncertainty of each parameter equal to the square root of the corresponding diagonal element of the covariance matrix. In our XRD analysis, we used an isotropic size-strain model, in which the lattice distortions associated with the non-hydrostatic stresses are primarily incorporated into the refined MS parameter. We refined all XRD patterns using the same procedure and instrumental parameters; therefore, the CS and MS values reported in the manuscript are reliable. The good agreement between the experimental and calculated profiles, together with the small residuals in the refined XRD patterns (Fig. S4), suggests that stress-related broadening is relatively weak and does not significantly affect our results. Moreover, stress-induced broadening is generally more significant in radial X-ray diffraction measurements, whereas its effect is considerably weaker in axial X-ray diffraction data. We followed the same procedures used in our previous axial X-ray diffraction studies[3,25,26]. We used E-WIMV and the March-Dollase texture models to fit the intensities of the BCC and HCP phases, respectively. As we used axial XRD, we observed a weak texture effect in our XRD patterns. We determined pressure in the BCC and HCP phases in runs 1 and 2 using equations of state obtained from hydrostatic experiments. We performed high-pressure XRD experiments on Fe-7%Mn at the 16-ID-B beamline, HPCAT, using a He pressure-transmitting medium and a ruby pressure calibrant, as described in detail in[30]. The estimated error in the pressure measurements is approximately 0.2 GPa, primarily due to uncertainties in determining the lattice parameter. We performed Digital Image Correlation (DIC) analysis of the sample surface during high-speed rotation of the piston-side anvil at 1,000 RPM for 120° using the open-source 2D subset-based digital image correlation software package Ncorr[48].

## Data availability

We will upload source data if the paper is accepted.

**Acknowledgements.** V. I. L acknowledges ARO (W911NF2420145), NSF (DMR-2246991 and CMMI-2519764), and Iowa State University (Murray Harpole Chair in Engineering). This work is performed at HPCAT (Sector 16), Advanced Photon Source (APS), and Argonne National Laboratory. HPCAT operations are supported by DOE-NNSA's Office of Experimental Science. The Advanced Photon Source is a U. S. Department of Energy (DOE) Office of Science User Facility operated for the DOE Office of Science by Argonne National Laboratory under Contract No. DE-AC02-06CH11357. We thank D. Popov and C. Kenney-Benson (HPCAT) for helping with laser-cutting the sample for run 3.

**Author Contributions.** S. Y. and M. S. performed experiments. S. Y. post-processed and analyzed the data, and M. S. assisted with post-processing and plotting. V. I. L. conceived the study, supervised the project, and developed theoretical analysis. S. Y and V. I. L analyzed the results. J. L. suggested and supplied the Fe-7%Mn alloy. S. S. designed and manufactured dRDAC, with the input and feedback from S. Y. and V. I. L. D. S. assisted with experiments at the APS and also performed real-time imaging of the sample at 1,000 RPM. S. Y. performed the digital image correlation analysis in discussion with V. I. L. V. I. L., S. Y, and M. S wrote the manuscript. A. S. performed FEM simulations for temperature calculation.

**Competing interests.** The authors declare no competing interests.

**Supplementary Information**

# Coupled plastic strain- and stress-induced phase transformations and microstructure evolution in Fe-7%Mn alloy in a dynamic rotational diamond anvil cell

Sorb Yesudhas[1]*, Mrinmay Sahu[1], Valery I. Levitas[1,2*], Dean Smith[3], Aniket Singh[1], Stanislav Sinogeikin[4], and Jeffrey T. Lloyd[5]

[1]Department of Aerospace Engineering, Iowa State University, Ames, Iowa 50011, USA

[2]Department of Mechanical Engineering, Iowa State University, Ames, Iowa 50011, USA

[3]HPCAT, X-ray Science Division, Argonne National Laboratory, Argonne, Illinois 60439, USA

[4]DAC Tools, LLC, Naperville, Illinois 60565, USA

[5]Armor Mechanisms Branch, DEVCOM Army Research Laboratory, Aberdeen Proving Ground, Maryland 21005, USA

*Corresponding authors. Email: sorbya@iastate.edu, vlevitas@iastate.edu

This file contains

**Supplementary Figures**

**Fig. S8:** Results of the digital image correlation analysis of the Fe-7%Mn sample at 1,000 RPM and 4.3 GPa at $r$ = 0 µm within dRDAC using a high-speed Photron camera (Run 3).

**Fig. S9:** Rotation speed and angle of an anvil vs. time in the experiment and simulations.

**Fig. S10:** Quarter anvil and sample assembly geometry and thermal boundary conditions used for axisymmetric transient thermal simulation.

**Fig. S11:** Evolution of radial temperature distribution in the sample.

**Fig. S12:** Results of FEM temperature simulations.

## Supplementary Tables

| | |
|---|---|
| Supplementary Table 1 | Pressures and unit-cell volumes of the BCC and HCP phases of Fe-7%Mn under compression and shear in run 1 at $r$ = 0 µm |
| Supplementary Table 2 | Pressures and unit-cell volumes of the BCC and HCP phases of Fe-7%Mn under compression and shear in run 1 at $r$ = 50 µm |
| Supplementary Table 3 | Pressures and unit-cell volumes of the BCC and HCP phases of Fe-7%Mn under compression and shear in run 2 at $r$ = 0 µm |
| Supplementary Table 4 | Pressures and unit-cell volumes of the BCC and HCP phases of Fe-7%Mn under compression and shear in run 2 at $r$ = 50 µm |
| Supplementary Table 5 | Time-resolved XRD study of Fe-7%Mn during torsion at an anvil speed of 100 RPM with 60° torsion at $r$ = 0 µm in run 1 |
| Supplementary Table 6 | Time-resolved XRD study of Fe-7%Mn during torsion at an anvil speed of 1,000 RPM with 180° torsion at $r$ = 110 µm in run 2 |
| Supplementary Table 7 | Time-resolved XRD study of Fe-7%Mn during torsion at an anvil speed of 1,500 RPM with 450° torsion at $r$ = 120 µm in run 2 |
| Supplementary Table 8 | Crystallite size (CS) and microstrain (MS) of the BCC and HCP phases of Fe-7%Mn at $r$ = 0 µm in run 1 |
| Supplementary Table 9 | Crystallite size (CS) and microstrain (MS) of Fe-7%Mn at $r$ = 50 µm in run 1 |
| Supplementary Table 10 | Crystallite size (CS) and microstrain (MS) of Fe-7%Mn at $r$ = 0 µm in run 2 |
| Supplementary Table 11 | Crystallite size (CS) and microstrain (MS) of Fe-7%Mn at $r$ = 50 µm in run 2 |
| Supplementary Table 12 | Calculation of thickness ($h$) and plastic strain ($\varepsilon_a$) at $r$ = 0 µm in runs 1 and 2 |
| Supplementary Table 13 | Calculation of axial plastic strain increment ($\Delta\varepsilon_a$), plastic strain rate ($\dot{\varepsilon_a}$), average shear strain increment ($\Delta\gamma_{avg}$), and maximum shear strain rate ($\dot{\gamma}_{max}$) |
| Supplementary Table 14 | Material properties used in thermal simulations |

**Supplementary Discussion**

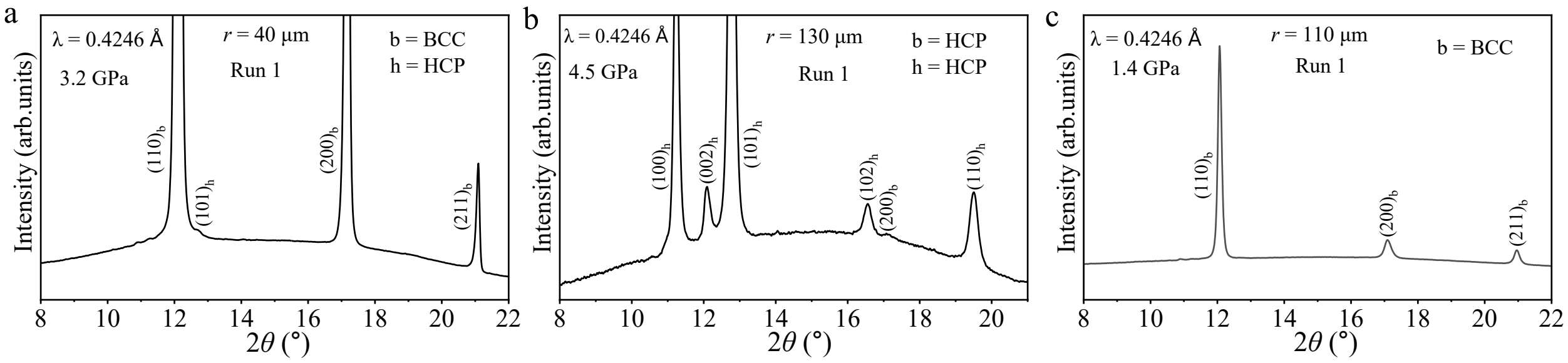

**Fig. S1**. **Direct and reverse PTs in Fe-7%Mn under compression and shear.** PT Initiation pressure of **(a)** BCC-HCP PT in run 1 (3.2 GPa at $r = 40$ μm and 3.5 GPa at $r = 0$ μm). **(b)** Initiation of HCP-BCC PT (4.5 GPa at $r = 130$ μm and 24.18 GPa at $r = 0$ μm). **(c)** Completion of the reverse HCP-BCC PT during unloading in run 1 (1.4 GPa at $r = 110$ μm and 24.18 GPa at $r = 0$ μm).

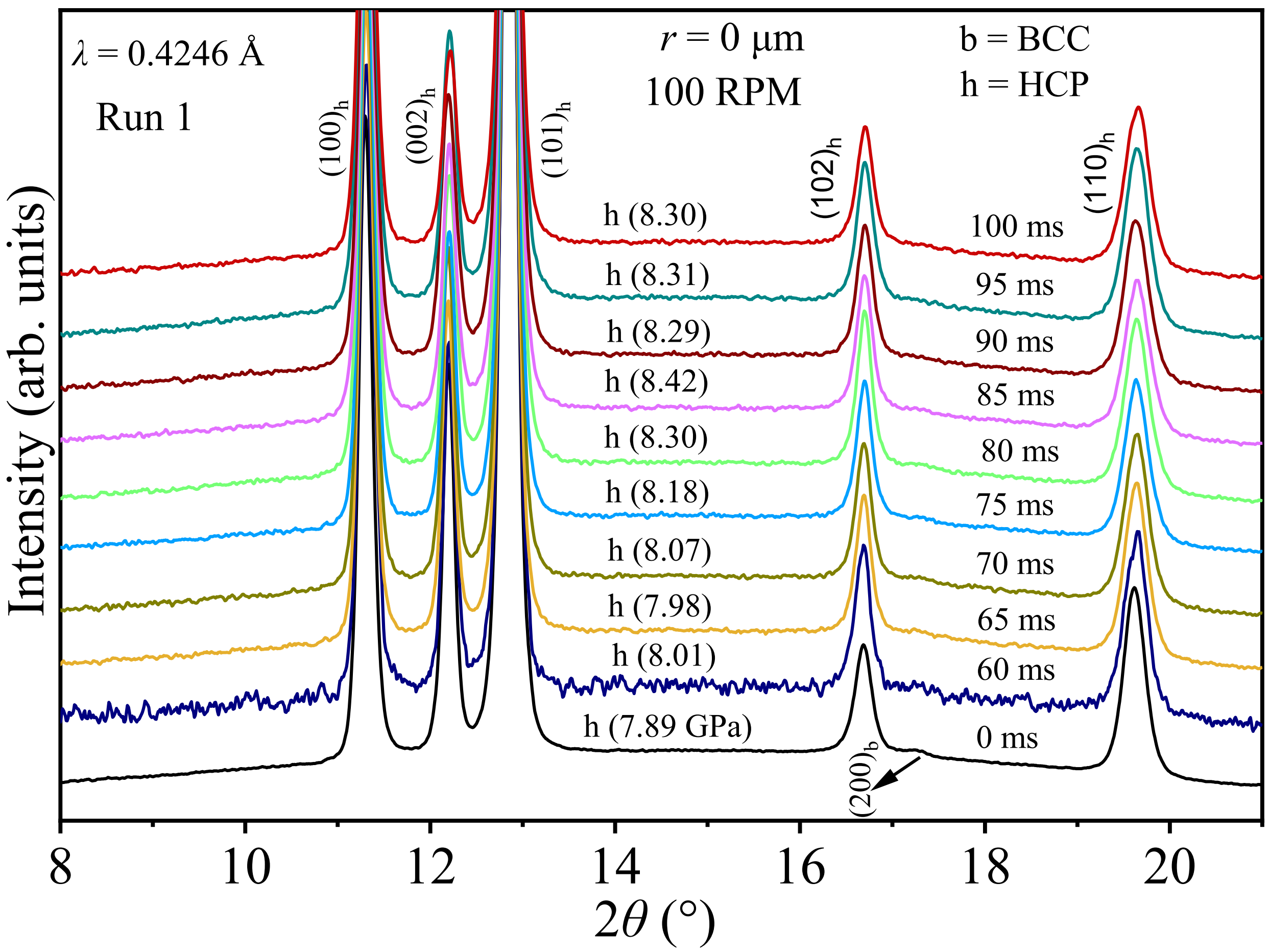

**Fig. S2. Real-time XRD study of Fe-7%Mn at 100 RPM**. XRD patterns of Fe-7%Mn at $r = 0$ μm and 100 RPM with 60° torsion in run 1. The pressure at the sample center ($r = 0$ μm) before torsion is 7.89 GPa. The 0 ms represents the XRD data before torsion. The data are available only from 60 ms onward. The BCC phase vanished after shear. $(200)_b$ is the peak from the BCC phase before the torsion (0 ms).

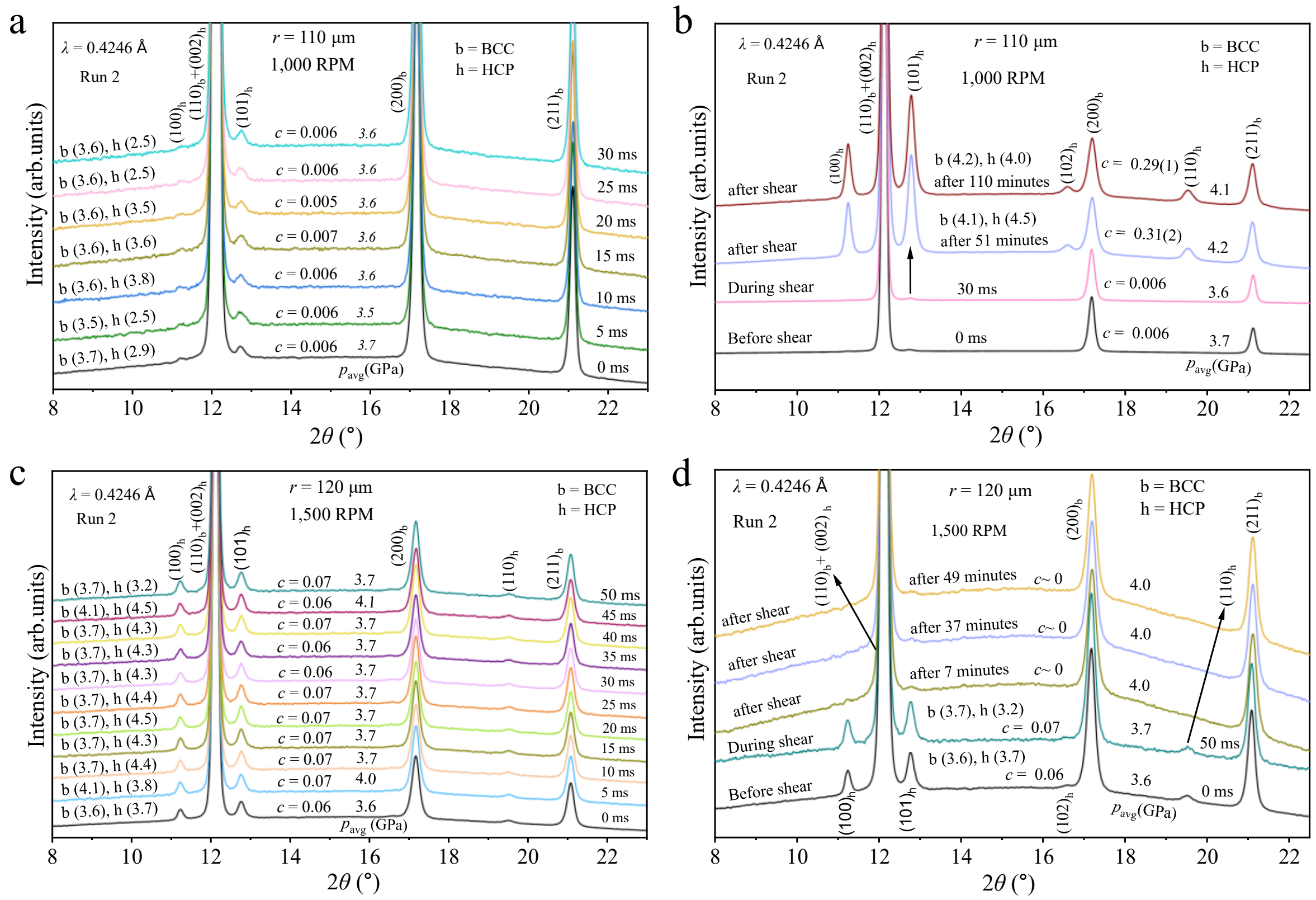

**Fig. S3. Real-time XRD study of Fe-7%Mn at 1,000 and 1,500 RPM.** XRD patterns of Fe-7%Mn **(a, b)** at $r = 110$ μm and 1,000 RPM with 180° torsion in run 2 and **(c, d)** at $r = 120$ μm and 1,500 RPM with 450° torsion in run 2. After torsion is completed, a dramatic increase in the volume fraction of the HCP phase in Fig. S3b and the disappearance of the HCP phase in Fig. S3d are shown. The uncertainties in the $c$ values provided in Fig. S3a are around 0.003. Similarly, the uncertainties in the $c$ values provided in Fig. S3c range from 0.005 to 0.011.

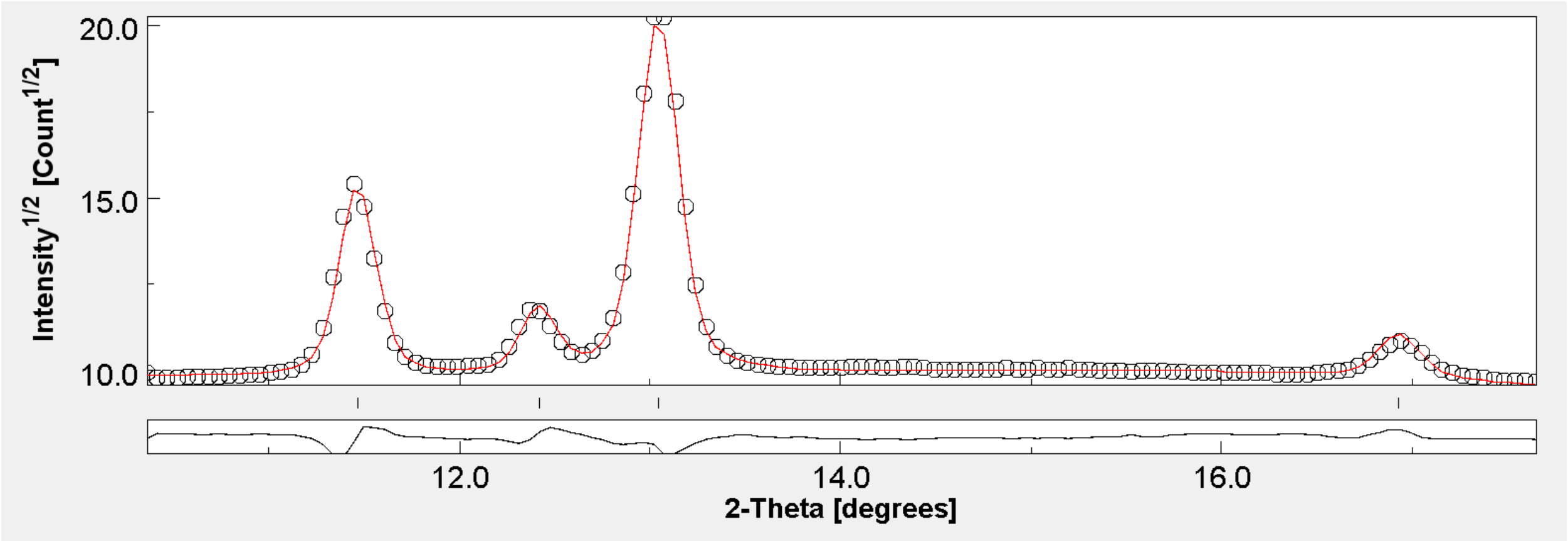

**Fig. S4: Rietveld refinement of the XRD pattern collected after $\varphi$ = 60° shear at $r$ = 50 µm and 100 RPM (Run 1).** Rietveld refinement of the azimuthally integrated XRD diffraction pattern of the HCP phase of Fe-7%Mn after slicing the XRD image into 72 slices. $p_{avg}$ = 6.11 GPa at $r$ = 50 µm (7.90 GPa at $r$ = 0 µm). The observed and calculated patterns are shown as circles and a red line, respectively. The difference plot is shown below the XRD pattern. The vertical marks represent the hkl Bragg positions.

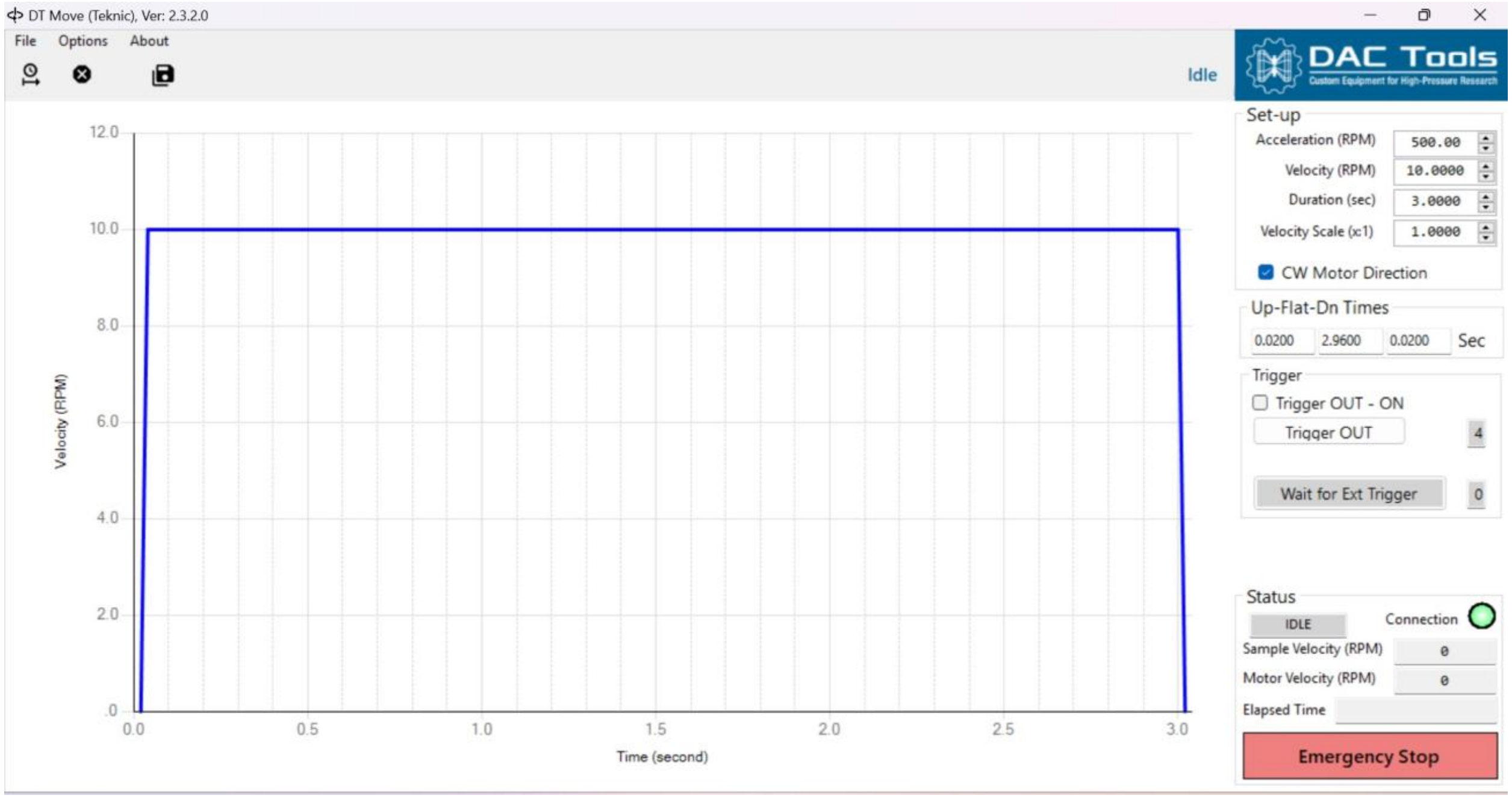

**Fig. S5. Velocity vs. Time profile in DT Move at 10 RPM.** The times for Ramp-up (0.02 s), Flat (2.96 s), and Ramp-down (0.02 s) are shown for the Up-Flat-Dn of the DT Move at 10 RPM.

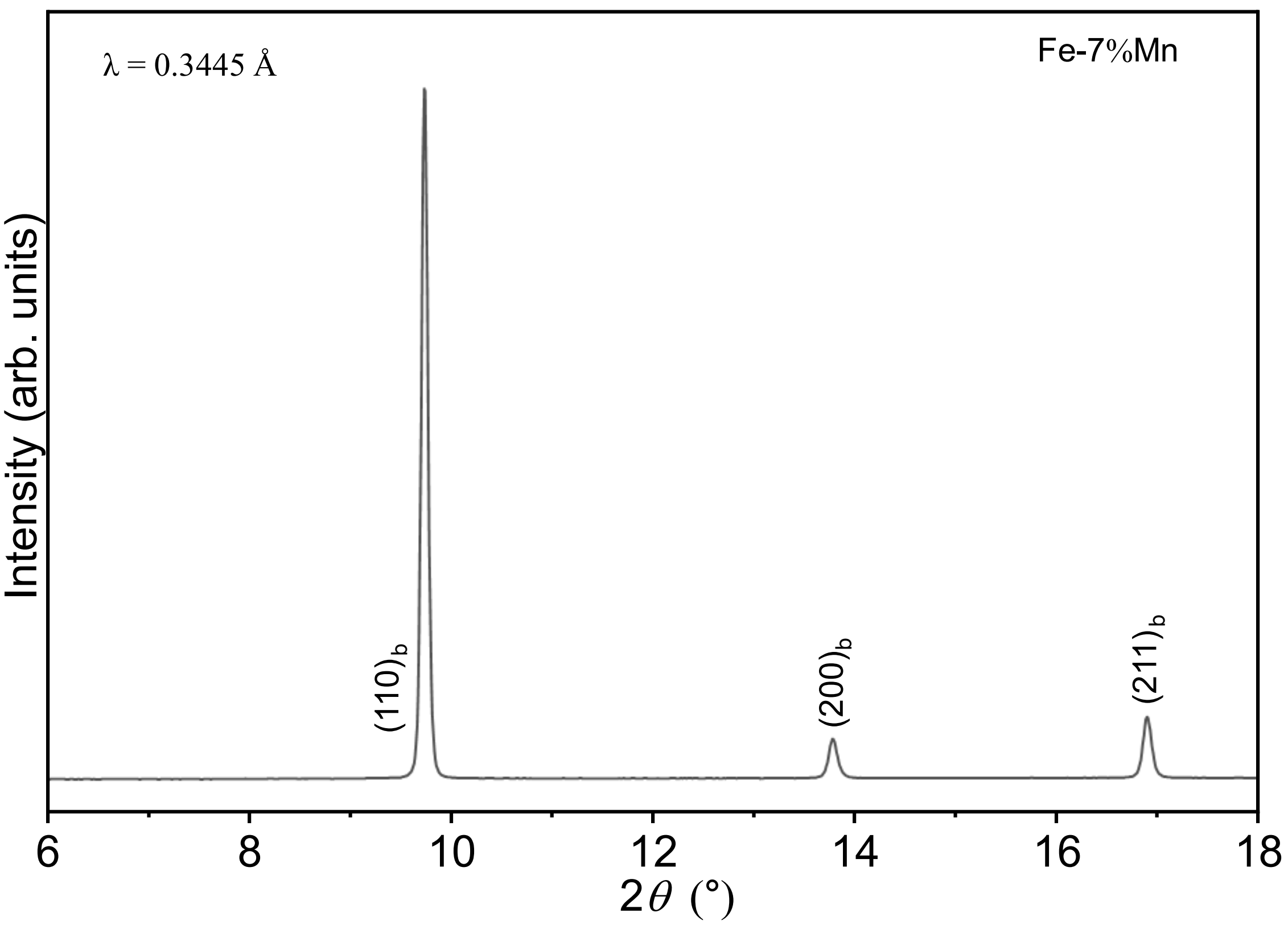

**Fig. S6. Ambient XRD pattern of Fe-7%Mn (Run 3).**

| $r$ = 0 (Run 1) | | | | | | |
|---|---|---|---|---|---|---|
| $p_{av}$ (GPa) | BCC $p$ (GPa) | HCP $p$ (GPa) | BCC $V$ (Å$^3$) | HCP $V$ (Å$^3$) | $c$ | Loading |
| 0.42 | 0.42 | | 23.524(48) | | 0 | Compression |
| 0.45 | 0.45 | | 23.519(40) | | 0 | Compression |
| 1.01 | 1.01 | | 23.438(38) | | 0 | Compression |
| 1.69 | 1.69 | | 23.342(36) | | 0 | Compression |
| 2.11 | 2.11 | | 23.284(38) | | 0 | Compression |
| 2.70 | 2.70 | | 23.202(46) | | 0 | Compression |
| 3.45 | 3.45 | 2.95 | 23.102(43) | 22.095(40) | 0.004(10) | Compression |
| 4.33 | 4.36 | 3.15 | 22.983(40) | 22.071(33) | 0.026(4) | Compression |
| 4.92 | 5.06 | 3.93 | 22.893(41) | 21.977(16) | 0.121(4) | Compression |
| 6.60 | 6.43 | 6.68 | 22.722(87) | 21.665(2) | 0.709(16) | 10 RPM, 30$^\circ$ |
| 7.89 | | 7.89 | | 21.534(2) | 1 | After 10 RPM, 180$^\circ$ |
| 7.92 | | 7.92 | | 21.531(1) | 1 | 100 RPM, 60$^\circ$ |
| 7.90 | | 7.90 | | 21.533(1) | 1 | After 100 RPM, 60$^\circ$ |
| 8.75 | | 8.75 | | 21.443(1) | 1 | Compression |
| 9.77 | | 9.77 | | 21.338(1) | 1 | Compression |
| 11.85 | | 11.85 | | 21.131(1) | 1 | Compression |
| 27.64 | | 27.64 | | 19.812(1) | 1 | After 100 RPM, 60$^\circ$ |
| 26.54 | | 26.54 | | 19.892(1) | 1 | After 100 RPM, 120$^\circ$ |

**Supplementary Table 1. Pressures and unit-cell volumes of the BCC and HCP phases of Fe-7%Mn under compression and shear in run 1 at $r$ = 0 µm**. $p_{av}$ denotes the average pressure over both phases, followed by the pressures in the BCC and HCP phases. The rightmost column lists the anvil loading details.

| $r$ = 50 (Run 1) | | | | | | |
|---|---|---|---|---|---|---|
| $p_{av}$ (GPa) | BCC $p$ (GPa) | HCP $p$ (GPa) | BCC $V$ (Å$^3$) | HCP $V$ (Å$^3$) | $c$ | Loading |
| 0.19 | 0.19 | | 23.557(38) | | 0 | Compression |
| 0.42 | 0.42 | | 23.523(36) | | 0 | Compression |
| 0.96 | 0.96 | | 23.446(37) | | 0 | Compression |
| 1.67 | 1.67 | | 23.345(35) | | 0 | Compression |
| 2.05 | 2.05 | | 23.293(35) | | 0 | Compression |
| 2.53 | 2.53 | | 23.226(35) | | 0 | Compression |
| 3.01 | 3.01 | | 23.161(38) | | 0 | Compression |
| 3.67 | 3.68 | 2.71 | 23.072(39) | 22.124(163) | 0.005(9) | Compression |
| 4.17 | 4.18 | 3.73 | 23.006(41) | 22.002(58) | 0.012(4) | Compression |
| 5.55 | 5.83 | 4.84 | 22.795(43) | 21.872(4) | 0.292(7) | 10 RPM, 40° |
| 5.85 | 5.84 | 5.85 | 22.794(68) | 21.757(2) | 0.744(23) | After 10 RPM, 180° |
| 6.11 | 5.85 | 6.19 | 22.793(58) | 21.720(1) | 0.768(33) | After 100 RPM, 60° |
| 7.02 | 6.16 | 7.31 | 22.755(59) | 21.596(1) | 0.751(19) | Compression |
| 7.99 | 7.49 | 8.12 | 22.593(59) | 21.509(1) | 0.789(34) | Compression |
| 9.83 | 7.93 | 9.97 | 22.541(141) | 21.317(1) | 0.930(31) | Compression |
| 19.00 | | 19.00 | | 20.483(1) | 1 | After 100 RPM, 60° |
| 18.94 | | 18.94 | | 20.488(1) | 1 | After 100 RPM, 120° |

**Supplementary Table 2. Pressures and unit-cell volumes of the BCC and HCP phases of Fe-7%Mn under compression and shear in run 1 at $r$ = 50 µm**. $p_{av}$ denotes the average pressure over both phases, followed by the pressures in the BCC and HCP phases. The rightmost column lists the anvil loading details.

.

| $r$ = 0 (Run 2) | | | | | | |
|---|---|---|---|---|---|---|
| $p_{av}$ (GPa) | BCC $p$ (GPa) | HCP $p$ (GPa) | BCC $V$ (Å$^3$) | HCP $V$ (Å$^3$) | $c$ | Loading |
| 0.94 | 0.94 | | 23.448(40) | | 0 | Compression |
| 2.03 | 2.03 | | 23.294(39) | | 0 | Compression |
| 2.71 | 2.71 | | 23.202(41) | | 0 | Compression |
| 3.19 | 3.20 | 2.02 | 23.136(39) | 22.209(90) | 0.0018(23) | Compression |
| 4.20 | 4.28 | 2.75 | 22.992(39) | 22.119(33) | 0.0518(35) | Compression |
| 4.95 | 5.30 | 4.25 | 22.862(60) | 21.940(13) | 0.3325(77) | Compression |
| 6.77 | 6.51 | 6.83 | 22.711(112) | 21.648(2) | 0.8042(382) | Compression |
| 9.52 | | 9.52 | | 21.364(2) | 1 | After 1,000 RPM, 180° |
| 9.71 | | 9.71 | | 21.343(1) | 1 | After 1,500 RPM, 450° |

**Supplementary Table 3. Pressures and unit-cell volumes of the BCC and HCP phases of Fe-7%Mn under compression and shear in run 2 at $r$ = 0 µm**. $p_{av}$ denotes the average pressure over both phases, followed by the pressures in the BCC and HCP phases. The rightmost column provides the anvil loading details.

| $r$ = 50 (Run 2) | | | | | | |
|---|---|---|---|---|---|---|
| $p_{av}$ (GPa) | BCC $p$ (GPa) | HCP $p$ (GPa) | BCC $V$ (Å$^3$) | HCP $V$ (Å$^3$) | $c$ | Loading |
| 0.66 | 0.66 | | 23.489(38) | | 0 | Compression |
| 1.75 | 1.75 | | 23.334(37) | | 0 | Compression |
| 2.51 | 2.51 | | 23.229(36) | | 0 | Compression |
| 2.98 | 2.98 | | 23.165(26) | | 0 | Compression |
| 3.83 | 3.84 | 1.12 | 23.050(36) | 22.322(13) | 0.0042(29) | Compression |
| 4.36 | 4.37 | 2.57 | 22.979(36) | 22.141(41) | 0.0159(29) | Compression |
| 5.63 | 5.90 | 5.08 | 22.787(52) | 21.844(5) | 0.3280(89) | Compression |
| 7.87 | | 7.87 | | 21.536(1) | 1 | after 1,000 RPM, 180° |
| 12.11 | | 12.11 | | 21.105(2) | 1 | after 1,500 RPM, 450° |

**Supplementary Table 4. Pressures and unit-cell volumes of the BCC and HCP phases of Fe-7%Mn under compression and shear in run 2 at *r* = 50 μm**. $p_{av}$ denotes the average pressure over both phases, followed by the pressures in the BCC and HCP phases. The rightmost column provides the anvil loading details.

| Real-time XRD study at $r$ = 0 μm and 100 RPM | | |
|---|---|---|
| Time (ms) | HCP $p$ (GPa) | HCP $V$ (Å$^3$) |
| 0 | 7.89 | 21.534(1) |
| 60 | 8.01 | 21.521(2) |
| 65 | 7.98 | 21.524(2) |
| 70 | 8.07 | 21.515(2) |
| 75 | 8.18 | 21.503(2) |
| 80 | 8.30 | 21.491(2) |
| 85 | 8.42 | 21.478(2) |
| 90 | 8.29 | 21.492(2) |
| 95 | 8.31 | 21.489(1) |
| 100 | 8.30 | 21.491(1) |

**Supplementary Table 5. Time-resolved XRD study of Fe-7%Mn during torsion at an anvil speed of 100 RPM with 60° torsion at $r$ = 0 μm in run 1.** We collected XRD patterns at $r$ = 0 μm with a data acquisition time of 5 ms. Time-resolved data were obtained from 60 ms onward. 0 ms represents data collected before torsion onset. The pressure in the HCP phase and the corresponding unit-cell volume evolution during torsion are listed in the second and third columns, respectively.

| Real-time XRD study at $r$ = 110 µm and 1,000 RPM | | | | | |
|---|---|---|---|---|---|
| Time (ms) | $p_{av}$ (GPa) | BCC $p$ (GPa) | HCP $p$ (GPa) | BCC $V$ (Å$^3$) | HCP $V$ (Å$^3$) |
| 0 | 3.65 | 3.65 | 2.93 | 23.0750(4) | 22.097(98) |
| 5 | 3.47 | 3.47 | 2.47 | 23.0990(3) | 22.154(72) |
| 10 | 3.60 | 3.60 | 3.81 | 23.0825(4) | 21.992(95) |
| 15 | 3.59 | 3.59 | 3.63 | 23.0833(4) | 22.013(84) |
| 20 | 3.59 | 3.59 | 3.46 | 23.0828(4) | 22.034(99) |
| 25 | 3.59 | 3.59 | 2.46 | 23.0822(4) | 22.154(102) |
| 30 | 3.57 | 3.57 | 2.49 | 23.0847(4) | 22.151(98) |
| After 51 minutes | 4.20 | 4.05 | 4.53 | 23.0223(4) | 21.908(4) |

**Supplementary Table 6. Time-resolved XRD study of Fe-7%Mn during torsion at an anvil speed of 1,000 RPM with 180° torsion at $r$ = 110 µm in run 2.** XRD patterns were collected at $r$ = 110 µm with a data acquisition time of 5 ms. 0 ms represents data collected before torsion onset. $p_{av}$ denotes the average pressure over both phases, followed by the pressures in the BCC and HCP phases. The last two columns show the evolution of the unit-cell volumes of the BCC and HCP phases during torsion.

| Real-time XRD study at $r$ = 120 µm and 1,500 RPM | | | | | |
|---|---|---|---|---|---|
| Time (ms) | $p_{av}$ (GPa) | BCC $p$ (GPa) | HCP $p$ (GPa) | BCC $V$ (Å$^3$) | HCP $V$ (Å$^3$) |
| 0 | 3.64 | 3.64 | 3.65 | 23.077(34) | 22.011(19) |
| 5 | 4.04 | 4.06 | 3.79 | 23.021(47) | 21.994(23) |
| 10 | 3.72 | 3.67 | 4.42 | 23.072(34) | 21.921(14) |
| 15 | 3.74 | 3.70 | 4.27 | 23.069(34) | 21.938(21) |
| 20 | 3.74 | 3.68 | 4.46 | 23.071(34) | 21.915(23) |
| 25 | 3.72 | 3.68 | 4.36 | 23.072(34) | 21.928(22) |
| 30 | 3.73 | 3.69 | 4.34 | 23.070(34) | 21.929(22) |
| 35 | 3.70 | 3.66 | 4.31 | 23.074(34) | 21.934(24) |
| 40 | 3.73 | 3.68 | 3.31 | 23.071(36) | 21.933(24) |
| 45 | 4.07 | 4.05 | 4.52 | 23.023(47) | 21.908(35) |
| 50 | 3.68 | 3.71 | 3.23 | 23.067(33) | 22.061(20) |
| After 7 minutes | 3.95 | 3.95 | | 23.036(47) | |
| After 37 minutes | 3.95 | 3.95 | | 23.036(47) | |
| After 47 minutes | 3.95 | 3.95 | | 23.036(47) | |

**Supplementary Table 7. Time-resolved XRD study of Fe-7%Mn during torsion at an anvil speed of 1,500 RPM with 450$^\circ$ torsion at $r$ = 120 µm in run 2**. We collected XRD patterns at $r$ = 120 µm with a data acquisition time of 5 ms. 0 ms represents data collected before torsion onset. $p_{av}$ denotes the average pressure over both phases, followed by the pressures in the BCC and HCP phases. The last two columns show the evolution of the unit-cell volumes of the BCC and HCP phases during torsion.

| $r = 0$ µm (Run 1) | | | | | | | | | |
|---|---|---|---|---|---|---|---|---|---|
| $p_{av}$ (GPa) | BCC CS (nm) | BCC *e* (nm) | BCC MS | BCC *e* (×10$^{-5}$) | HCP CS (nm) | HCP *e* (nm) | HCP MS | BCC *e* (×10$^{-5}$) | Loading |
| 0.42 | 337.06 | 57.57 | 0.00221 | 7.67 | | | | | Compression |
| 0.45 | 329.57 | 51.17 | 0.00219 | 6.18 | | | | | Compression |
| 1.01 | 317.61 | 46.02 | 0.00239 | 5.51 | | | | | Compression |
| 1.69 | 289.58 | 45.33 | 0.00243 | 5.33 | | | | | Compression |
| 2.11 | 251.00 | 32.57 | 0.00248 | 5.55 | | | | | Compression |
| 2.70 | 224.87 | 24.43 | 0.0025 | 6.74 | | | | | Compression |
| 3.45 | 220.43 | 28.73 | 0.00251 | 6.58 | | | | | Compression |
| 4.33 | 217.83 | 26.33 | 0.00247 | 6.05 | 29.95 | 22.63 | 0.00374 | 101.00 | Compression |
| 4.92 | 212.73 | 26.11 | 0.00258 | 6.06 | 28.20 | 6.39 | 0.00398 | 80.20 | Compression |
| 6.60 | 100.12 | 13.32 | 0.00363 | 1.24 | 27.75 | 0.83 | 0.00328 | 6.97 | 10 RPM, 30° |
| 7.89 | | | | | 27.16 | 0.49 | 0.00351 | 9.25 | After 10 RPM, 180° |
| 7.92 | | | | | 27.02 | 0.54 | 0.00319 | 9.83 | 100 RPM, 60° |
| 7.90 | | | | | 27.24 | 0.53 | 0.00322 | 9.31 | After 100 RPM, 60° |
| 8.75 | | | | | 27.16 | 0.56 | 0.00308 | 10.20 | Compression |
| 9.77 | | | | | 26.87 | 0.52 | 0.00316 | 10.00 | Compression |
| 11.85 | | | | | 26.98 | 0.61 | 0.00327 | 10.60 | Compression |
| 27.64 | | | | | 26.96 | 0.67 | 0.00290 | 12.30 | After 100 RPM, 60° |
| 26.54 | | | | | 26.67 | 0.68 | 0.00286 | 11.80 | After 100 RPM, 120° |

**Supplementary Table 8**: **Crystallite size (CS) and microstrain (MS) of the BCC and HCP phases of Fe-7%Mn at *r* = 0 µm in run 1.** CS and MS with error (*e*) of Fe-7%Mn under high pressure and shear at $r = 0$ µm (run 1). $p_{av}$ denotes the average pressure over both phases.

| $r$ = 50 µm (Run 1) | | | | | | | | | |
|---|---|---|---|---|---|---|---|---|---|
| $p_{av}$ (GPa) | BCC CS (nm) | BCC $e$ (nm) | BCC MS | BCC $e$ ($\times10^{-5}$) | HCP CS (nm) | HCP $e$ (nm) | HCP MS | BCC $e$ ($\times10^{-5}$) | Loading |
| 0.19 | 356.74 | 57.23 | 0.00217 | 5.76 | | | | | Compression |
| 0.42 | 339.15 | 49.20 | 0.00224 | 5.48 | | | | | Compression |
| 0.96 | 329.02 | 52.58 | 0.00244 | 3.62 | | | | | Compression |
| 1.67 | 286.85 | 34.09 | 0.00258 | 3.43 | | | | | Compression |
| 2.05 | 265.9 | 29.66 | 0.00270 | 4.86 | | | | | Compression |
| 2.53 | 249.26 | 26.48 | 0.00278 | 4.84 | | | | | Compression |
| 3.01 | 237.5 | 26.56 | 0.00285 | 4.08 | | | | | Compression |
| 3.67 | 243.66 | 32.69 | 0.00287 | 5.46 | | | | | Compression |
| 4.17 | 226.66 | 29.31 | 0.00289 | 5.71 | 30.16 | 20.82 | 0.00403 | 196.00 | Compression |
| 5.55 | 208.55 | 31.32 | 0.00389 | 5.58 | 27.83 | 1.45 | 0.00358 | 21.10 | 10 RPM, 40° |
| 5.85 | 155.22 | 27.73 | 0.00377 | 9.15 | 27.29 | 1.00 | 0.00352 | 12.40 | After 10 RPM, 180° |
| 6.11 | 160.31 | 26.45 | 0.00361 | 8.44 | 27.06 | 0.58 | 0.00318 | 10.20 | After 100 RPM, 60° |
| 7.02 | 122.09 | 13.51 | 0.00336 | 8.46 | 26.78 | 0.44 | 0.00282 | 16.60 | Compression |
| 7.99 | 116.15 | 12.21 | 0.00326 | 9.16 | 26.40 | 0.52 | 0.00287 | 9.85 | Compression |
| 9.83 | 107.78 | 22.78 | 0.00292 | 20.6 | 26.41 | 0.42 | 0.00292 | 8.45 | Compression |
| 19.00 | | | | | 25.66 | 0.48 | 0.00410 | 7.30 | After 100 RPM, 60° |
| 18.94 | | | | | 24.78 | 0.41 | 0.00393 | 7.26 | After 100 RPM, 120° |

**Supplementary Table 9**: **Crystallite size (CS) and microstrain (MS) of Fe-7%Mn at *r* = 50 µm in run 1.** CS and MS with error (*e*) of Fe-7%Mn under high pressure and shear at *r* = 50 µm (run 1). $p_{av}$ denotes the average pressure over both phases.

| $r$ = 0 µm (Run 2) | | | | | | | | | |
|---|---|---|---|---|---|---|---|---|---|
| $p_{av}$ (GPa) | BCC CS (nm) | $e$ (nm) | BCC MS | $e$ (×10$^{-5}$) | HCP CS (nm) | $e$ (nm) | HCP MS | $e$ (×10$^{-4}$) | Loading |
| 0.94 | 279.03 | 37.70 | 0.00236 | 5.96 | | | | | Compression |
| 2.03 | 265.77 | 41.26 | 0.00245 | 5.78 | | | | | Compression |
| 2.71 | 237.53 | 32.38 | 0.00245 | 6.16 | | | | | Compression |
| 3.19 | 230.06 | 23.08 | 0.00249 | 5.65 | | | | | Compression |
| 4.20 | 221.09 | 25.96 | 0.00255 | 5.74 | 28.04 | 12.32 | 0.00376 | 8.74 | Compression |
| 4.95 | 193.48 | 32.59 | 0.00275 | 8.93 | 27.57 | 5.47 | 0.00428 | 6.91 | Compression |
| 6.77 | 151.45 | 33.43 | 0.00326 | 16.30 | 26.82 | 0.86 | 0.00232 | 2.25 | Compression |
| 9.52 | | | | | 26.11 | 0.62 | 0.00224 | 0.53 | after 1,000 RPM, 180° |
| 9.71 | | | | | 25.96 | 0.42 | 0.00228 | 1.00 | after 1,500 RPM, 450° |

**Supplementary Table 10**: **Crystallite size (CS) and microstrain (MS) of Fe-7%Mn at *r* = 0 µm in run 2.** CS and MS with error (*e*) of Fe-7%Mn under high pressure and shear at *r* = 0 µm (run 2). $p_{av}$ denotes the average pressure over both phases.

| $r$ = 50 µm (Run 2) | | | | | | | | | |
|---|---|---|---|---|---|---|---|---|---|
| $p_{av}$ (GPa) | BCC CS (nm) | $e$ (nm) | BCC MS | $e$ (×10$^{-5}$) | HCP CS (nm) | $e$ (nm) | HCP MS | $e$ (×10$^{-4}$) | Loading |
| 0.66 | 257.85 | 29.20 | 0.00242 | 5.54 | | | | | Compression |
| 1.75 | 235.58 | 25.15 | 0.00267 | 5.17 | | | | | Compression |
| 2.51 | 217.40 | 23.87 | 0.00275 | 5.14 | | | | | Compression |
| 2.98 | 205.54 | 14.12 | 0.00276 | 3.69 | | | | | Compression |
| 3.83 | 195.19 | 18.94 | 0.00282 | 5.09 | | | | | Compression |
| 4.36 | 186.37 | 16.69 | 0.00282 | 5.04 | 29.63 | 2.06 | 0.00278 | 4.02 | Compression |
| 5.63 | 168.66 | 17.52 | 0.00318 | 7.12 | 28.26 | 1.70 | 0.00309 | 3.21 | Compression |
| 7.87 | | | | | 27.10 | 0.49 | 0.00348 | 1.00 | After 1,000 RPM, 180° |
| 12.11 | | | | | 26.11 | 0.61 | 0.00398 | 1.16 | After 1,500 RPM, 450° |

**Supplementary Table 11**: **Crystallite size (CS) and microstrain (MS) of Fe-7%Mn at *r* = 50 µm in run 2.** CS and MS with error ($e$) of Fe-7%Mn under high pressure and shear at $r$ = 50 µm (run 2). $p_{av}$ denotes the average pressure over both phases.

| Run 1 ($r$ = 0 µm) | | | | Run 2 ($r$ = 0 µm) | | | |
|---|---|---|---|---|---|---|---|
| $P_{avg}$ (GPa) | $h$ (µm) | $\varepsilon_a$ | Loading | $P_{avg}$ (GPa) | $h$ (µm) | $\varepsilon_a$ | Loading |
| $10^{-4}$ | 168.0 | 0.00 | Compression | $10^{-4}$ | 159.0 | 0.00 | Compression |
| 0.42 | 164.6 | 0.02 | Compression | 0.20 | 158.7 | 0.002 | Compression |
| 0.45 | 164.6 | 0.02 | Compression | 0.50 | 157.1 | 0.01 | Compression |
| 1.01 | 163.8 | 0.03 | Compression | 0.94 | 153.4 | 0.04 | Compression |
| 1.69 | 158.4 | 0.06 | Compression | 2.03 | 135.6 | 0.16 | Compression |
| 2.11 | 150.2 | 0.11 | Compression | 2.71 | 118.0 | 0.30 | Compression |
| 2.70 | 136.9 | 0.21 | Compression | 3.19 | 110.2 | 0.37 | Compression |
| 3.45 | 125.2 | 0.29 | Compression | 4.20 | 95.8 | 0.51 | Compression |
| 4.33 | 114.2 | 0.39 | Compression | 6.77 | 75.3 | 2.11 | Compression |
| 4.92 | 108.7 | 0.44 | Compression | 9.52 | 18.0 | 2.18 | After 1,000 RPM, 180° |
| 7.89 | 60.3 | 1.03 | After 10 RPM, 180° | 9.71 | 8.2 | 2.97 | After 1,500 RPM, 450° |
| 7.90 | 19.4 | 2.16 | After 100 RPM, 60° | | | | |
| 8.75 | 18.9 | 2.19 | Compression | | | | |
| 11.85 | 20.4 | 2.11 | Compression | | | | |
| 27.64 | 13.4 | 2.53 | After 100 RPM, 60° | | | | |
| 26.54 | 11.8 | 2.66 | After 100 RPM, 120° | | | | |

**Supplementary Table 12: Calculation of thickness (*h*) and plastic strain ($\varepsilon_a$) at *r* = 0 µm in runs 1 and 2.** $p_{av}$ denotes the average pressure over both phases. $\varepsilon_a$ is calculated from the initial ($h_0$) and final ($h_f$) thicknesses using the relation $\varepsilon_a = \ln(\frac{h_0}{h_f})$.

| Anvil speed (RPM), $\varphi$ (°) | $p_{avg}$ (GPa) | $h_i$ (µm) | $h_f$ (µm) | $h_{avg}$ (µm) | $r$ (µm) | $\Delta\varepsilon_a$ ($r$ = 0 µm) | $\dot{\varepsilon_a} = \frac{\Delta\varepsilon_a}{\Delta t}$ (s$^{-1}$) | $\Delta\gamma_{avg}$ | $\dot{\gamma}_{max}$ (s$^{-1}$) |
|---|---|---|---|---|---|---|---|---|---|
| 10, 180° | 7.89 | 108.7 | 60.3 | 84.5 | 120 | 0.59 | 0.20 | 4.46 | 2.08 |
| 100, 60° | 8.30 | 60.3 | 19.4 | 39.9 | 0 | 1.13 | 11.30 | 0 | 0 |
| 1,000, 180° | 3.57 | 75.3 | 18.0 | 46.7 | 110 | 1.43 | 47.70 | 7.40 | 639.95 |
| 1,500, 450° | 3.68 | 18.0 | 8.2 | 13.1 | 120 | 0.79 | 15.8 | 71.94 | 2,298.73 |

**Supplementary Table 13: Calculation of axial plastic strain increment ($\Delta\varepsilon_a$), plastic strain rate ($\dot{\varepsilon_a}$), average shear strain increment ($\Delta\gamma_{avg}$), and maximum shear strain rate ($\dot{\gamma}_{max}$).** For $\dot{\varepsilon_a}$, $\Delta t$ = 3 s for 10 RPM, 100 ms for 100 RPM, 30 ms for 1,000 RPM, and 50 ms for 1,500 RPM. $\Delta\varepsilon_a$ is calculated from the initial ($h_i$) and final ($h_f$) thicknesses for each loading increment using the relation $\Delta\varepsilon_a = \ln(\frac{h_i}{h_f})$. $\dot{\gamma}_{max}$ is evaluated for the final thickness $h_f$. $p_{av}$ denotes the average pressure over both phases. The pressures for 100, 1,000, and 1,500 RPM correspond to the XRD data collected at 100, 30, and 50 ms, respectively.

**Supplementary Discussion**

## I. Dynamic Rotational Diamond Anvil Cell (dRDAC)

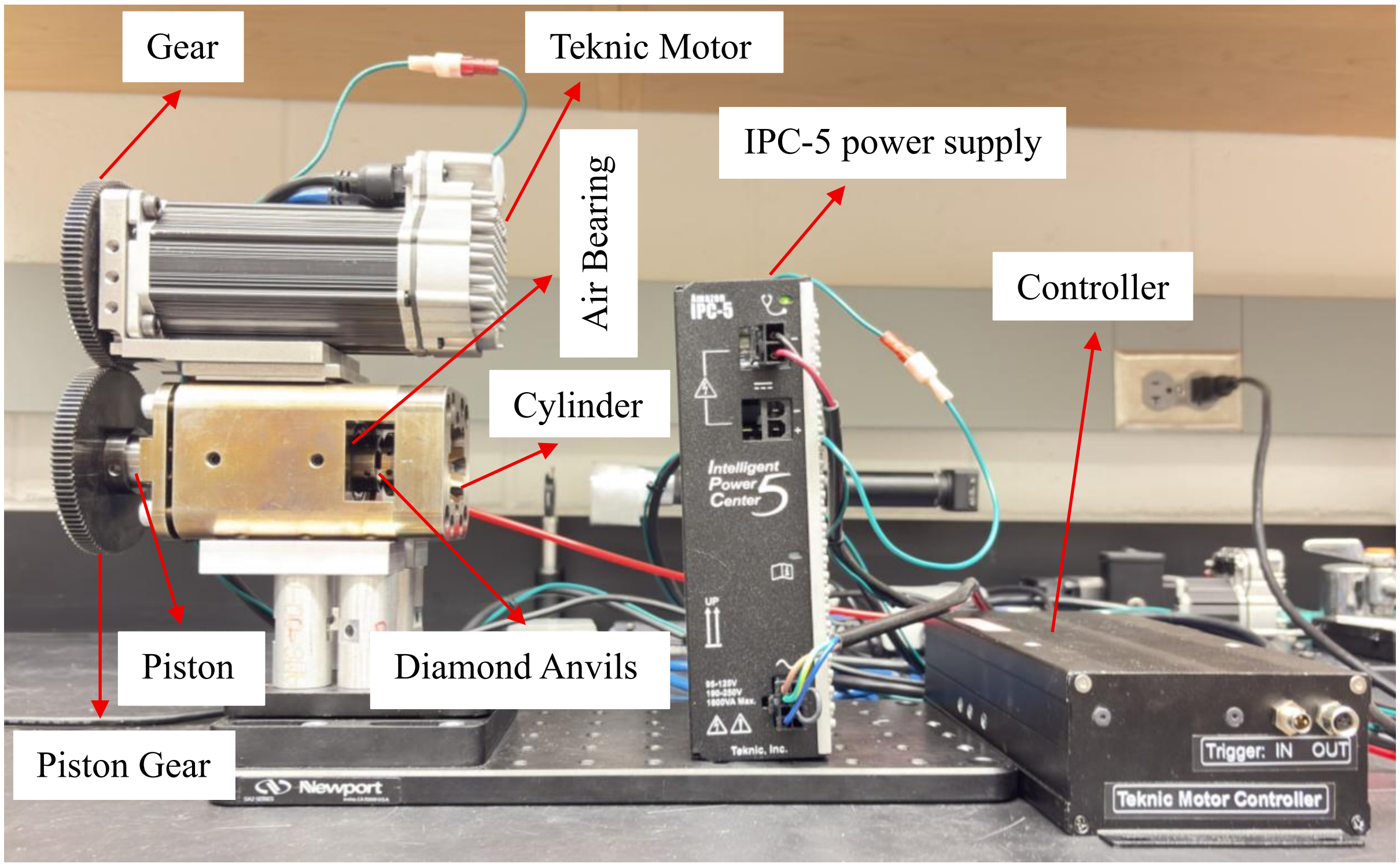

**Fig. S7. Components of the dRDAC setup.**

The dynamic rotational diamond anvil cell (dRDAC) is an advanced version of the rotational diamond anvil cell (RDAC) designed to achieve high strain rates of up to $10^4$ $s^{-1}$. The system integrates a piston-cylinder assembly, a graphite air bearing, a gear transmission mechanism, a Teknic motor, an IPC-5 power supply, an SC4-HUB controller powered by a 24 VDC supply, and DT Move software for precise control of the rotational motion (Fig. S7). The piston rotates freely inside a graphite air bearing operating at 80-90 psi, greatly reducing friction between the piston and the bearing. Shear deformation is generated by coupling the gear attached to the dRDAC piston with the gear mounted on the rotor shaft of the Teknic motor. The Teknic SC motor is powered by an IPC-5 power supply and controlled through the DAC Tools controller, which consists of a Teknic SC4-HUB and a 24 VDC power supply (Mean Well LRS-150-24). The 24 VDC supply powers the electronic components of the SC4-HUB, which serves as the communication interface between the computer, input/output devices, and the Teknic SC motor[1]. Although the motor can be operated using Teknic ClearView software, this study used the Visual Studio-based DT Move software developed by DAC Tools, LLC. This software enables real-time control and monitoring of the motor parameters, including acceleration, velocity, and duration, as illustrated in Fig. S5. This program also allows an external signal to trigger the motor.

## II. Calculation of plastic and shear strains

Because one cannot follow the straining path of each particle inside the sample during the experiment, we use the following approach to calculate the plastic axial and shear strains.

The axial plastic strain increment is

$$\Delta\varepsilon_a = \ln\left(\frac{h_i}{h_f}\right), \qquad \text{(S1)}$$

where $h_i$ and $h_f$ are the initial and final thicknesses for the given deformation increment.

When torsion is applied at a constant thickness, the shear strain increment during a change in the angle of $\Delta\varphi$ is

$$\Delta\gamma = \frac{\Delta\varphi\, r}{h}, \qquad \text{(S2)}$$

where $r$ is the distance from the sample center ($r = 0$) in μm, and $h$ is the sample thickness in μm.

Since thickness varies during the torsion from $h_i$ to $h_f$, the average thickness is used for the shear strain increment calculation as follows.

$$\Delta\gamma = \frac{\Delta\varphi\, r}{0.5(h_i+h_f)}. \qquad \text{(S3)}$$

Therefore, an increment in accumulated plastic strain is

$$\Delta q = \frac{\Delta\gamma}{\sqrt{3}} = \frac{2\,\Delta\varphi\, r}{\sqrt{3}(h_i+h_f)}. \qquad \text{(S4)}$$

The total axial, shear, and accumulated plastic strain is the sum of contributions at each loading step.

Shear strain rate for a given thickness is,

$$\dot{\gamma} = \frac{\omega\, r}{h} \qquad \text{(S5)}$$

where $\omega = 2\pi N/60$ rad/s is the angular velocity, and $N$ is the anvil speed in RPM.

**Run 1**

**For 10 RPM**

For $\omega$ = 10 RPM, $\Delta\varphi = 180^\circ = 180\times\pi/180 = \pi$, $h_i$ = 108.7 μm, $h_f$ = 60.3 μm, $h_{avg}$ = 84.5 μm, $r$ = 120 μm,

Shear strain increment $\Delta\gamma = \frac{\pi\times120}{(108.7+60.3)/2} = 4.46$;

Plastic strain increment $\Delta q = \frac{4.46}{\sqrt{3}} = 2.58$;Maximum shear strain rate for the minimal thickness $h_f$

$$\dot{\gamma}_{max} = \frac{r\times\omega}{h_f} = \frac{120\times(2\times\pi\times\frac{10}{60})}{60.3} = 2.08\ s^{-1}$$

**Run 2**

**For 1,000 RPM**

For $\omega$ = 1,000 RPM, $\Delta\varphi = 180^\circ = \pi$, $h_i$ = 75.3 μm, $h_f$ = 18 μm, $h_{avg}$ = 46.7 μm, $r$ = 110 μm

Shear strain increment $\Delta\gamma = \frac{\pi\times110}{0.5(75.3+18.0)} = 7.40$;

Plastic strain increment $\Delta q = \frac{7.40}{\sqrt{3}} = 4.27$.

Maximum shear strain rate, $\dot{\gamma}_{max} = \frac{r\times\omega}{h_f} = \frac{(2\times\pi\times\frac{1,000}{60})\times110}{18} = 639.95\ s^{-1}$.

**For 1500 RPM**

For $\omega$ = 1,500 RPM, $\Delta\varphi = 450^\circ = 450\times\pi/180$, $h_i$ = 18 μm, $h_f$ = 8.2 μm, $r$ = 120 μm

$$\Delta\gamma = \frac{450\times\frac{\pi}{180}\times120}{0.5(18.0+8.2)} = 71.94;$$

Accumulated plastic strain increment $\Delta q = \frac{71.94}{\sqrt{3}} = 41.53$.

Maximum shear strain rate $\dot{\gamma}_{max} = \frac{r\times\omega}{h_f} = \frac{(2\times\pi\times\frac{1,500}{60})\times120}{8.2} = 2,298.73\ s^{-1}$ (for final thickness).

## III. Strategy and planning of the experimental program

For experiments in dRDAC during rotation up to 10 RPM, we can scan the beam along the sample diameter to collect real-time XRD data. For 10 RPM experiments with a collection time of 10 ms between two closest data points, the rotation angle increment is just $2^\circ$. During rotation in fast dRDAC for ≥ 100 RPM, real-time XRD data at the fixed load can be collected at a single spatial point of a sample only because of the large rotation increment between the two nearest points.

Preparation of the experiments in dRDAC with rough diamond anvils is much more time-consuming than in traditional RDAC and with smooth anvils. This is because the piston in the dynamic RDAC is moving within a graphite bearing at an air pressure of 80-90 psi. After continuous use of the dRDAC, graphite powder particles are gradually removed due to the friction between the graphite and the piston. Therefore, it takes between two weeks and one month to align the dRDAC at the current stage. After a single experimental program in dRDAC at high strain rates and load release, diamonds are very often laterally displaced after large rotations and require realignment. Consequently, only one dRDAC experiment is possible per beamtime. Therefore, a very limited number of experiments can be performed, and a systematic parametric study is questionable. For example, experiments for the same $n$ pressures and sample thicknesses (which vary significantly during torsion, likely differently for different rotation rates) and $m$ different rotation rates require m·n beamtimes, which is unrealistic until a faster alignment procedure is developed.

Also, finite element method solutions[15,23,24,36,37] show strong heterogeneity in all components of the stress and plastic strain tensors, accumulated plastic strain, and volume fraction of the high-pressure phase fields along both radius and thickness. Therefore, the equation for the average over the thickness shear strain increment $\Delta\gamma = \frac{\Delta\varphi\, r}{h}$, used in SPD literature[1,2,4,6,7] gives a rough estimate. Most importantly, kinetic rules must be formulated for a *material point*. During torsion, significant thickness reduction and radial plastic flow occur. Consequently, the XRD beam collects data from material points from different initial radii (and, consequently, pressure-plastic strain paths) that pass through the selected spatial point.

Therefore, combined computational-experimental methods were suggested[23,24], in which a developed material model and corresponding finite element solutions were fitted to the obtained limited experimental data, resulting in the determination of all material parameters and the distribution of the stress and plastic strain tensors, accumulated plastic strain, and volume fraction of the high-pressure phase fields. This complex, long, iterative process has so far been implemented only for compression in DAC.

With these challenges and perspectives in mind, and given that these are the first experiments on strain-induced PTs in Fe-7%Mn and in dRDAC for any material, our goal is to perform two trial experiments with various combinations of pressures, rotation angles, speeds, and radii. A wide range of control parameters helps discover new phenomena and rules, as shown in our previous publications[3,13-15,23-27]. This allowed us to find here a steady-state solution to the strain-induced PT kinetics and an important rule for microstructure evolution in the HCP phase experimentally.

Our plan was based on traditional wisdom that time is not an important parameter (because in dynamic experiments, PTs occur even at nano- and picosecond time scales, and plastic straining can substitute for increased temperature) and that, if PT starts at the given pressure, it can be completed by increasing plastic deformation. Both assumptions proved wrong, which led to the main discoveries of this work but required adjusting the experimental plan during the experiments.

In Run 1, we performed static compression up to the initiation of PT ($c$ = 0.121(4)) at the center of the sample, then scanned along the sample diameter. During each scan along the diameter, we collected XRD data for 30 points. For 10 RPM and torsion by 180$^{\circ}$ (a relatively large angle to understand the effect of shear at relatively low pressures), continuous scanning along the diameter was performed during torsion. Since the HCP phase is much stronger than the BCC phase, we expected pressure to increase at the center, allowing it to cover a broad pressure range, which was indeed the case.

After completing both experiments and working on the paper, we decided not to present distributions of pressure, $c$, and microstructural parameters along the diameter, which were measured at 30 points. This would not add to the main results summarized in the abstract and conclusions. Still, it would require extensive post-processing and analysis, and we will present it in a separate paper with different goals, along with detailed finite element analysis and additional experiments. Instead, we present results for $r$ = 0, where shear is zero and pressure is maximum, and for $r$ = 50 μm, where shear strain and strain rate are relatively low and the PT completes. We also presented selected results for $r$ = 40, 130, and 110 μm (Fig. S1), where minimum pressures for the initiation of the direct PT $p_{\varepsilon}^{d} = 3.20$ GPa, and the maximum pressure for initiating (4.5 GPa) and completing (1.4 GPa) the reverse PT during unloading were found.

For real-time measurements at 100 RPM, we selected $r$ = 0, where shear is zero and pressure is at its maximum, to study PT and microstructure evolution under a controlled strain rate during torsion-induced compression. For real-time measurements at high rotation rates, we selected $r \leq$ 120 μm, where the anvil-edge effect is negligible, to achieve the highest strain rate. Thus, for the maximum 1,500 RPM, we chose $r$ = 120 μm. Similarly, for real-time measurements at the lower rotation rate of 1,000 RPM, we selected $r$ = 110 μm, where PT had just started. We are equally interested in PT and microstructure evolution.

After completing torsion at 10 RPM, we collected data along the sample diameter both during and after shear. After completing shear, the BCC phase was almost transformed to the HCP phase. Then, we performed torsion at 100 RPM for 60°. We were surprised that the pressure (7.9 GPa) and the volume fraction of the HCP phase did not change significantly at $r$ = 50 μm, indicating an approach to steady state, which we did not expect. Therefore, we increased the pressure to 11.9 GPa at the center and rotated by another 60°, resulting in an unexpected, drastic increase to 27.6 GPa. The final 120° rotation did not change the pressure at the sample center. We did not continue loading because the BCC phase completely transformed into the HCP phase throughout the entire sample region; only a small volume fraction of the BCC phase was observed near the sample edge, the microstructure was steady, and the thickness was already quite small (11.8 μm), which could lead to damage to the diamonds during further rotation and further thickness reduction.

For run 2, we compressed the sample in several steps, during which we confirmed $p_{\varepsilon}^{d} = 3.2$ GPa at the center (Fig. 2a) until the PT started ($c$ = 0.006(27)) at $p$ = 3.65 GPa > $p_{\varepsilon}^{d}$) at $r$ = 110 μm (Supplementary Table 6). Then we twisted at 1,000 RPM by 180°, enough to collect 7 real-time points at a constant rotation speed. During this time, we analyzed the XRD patterns and were shocked to find that PT did not occur. After the torsion stopped, we performed the sample centering in the Diptera software. However, due to some technical difficulties, we collected data only after 51 minutes along the sample diameter. After 51 minutes, the data collected at $r$ = 110 μm showed that the volume fraction had increased to 0.31(2) and remained constant thereafter (Fig. S3b). Next, we performed torsion at 1,500 RPM and measured at $r$ = 120 μm, where $p$ = 3.64 GPa and $c$ = 0.06(1). After completing shear, we collected data along the sample diameter after 7 minutes (Fig. S3c and Supplementary Table 7). We observed that the reverse PT was completed and that no further changes occurred. Because this was the first experiment at an anvil speed of 1,500 RPM, we did not continue due to uncertainty about the diamond's condition and the small sample thickness of 8.2 μm. We performed no further experiments because the allocated beamtime ended.

In addition to the results reported in the paper, these experiments helped us significantly correct and refine the plan for future experiments.

## IV. Absence of contact sliding between the sample and the diamond

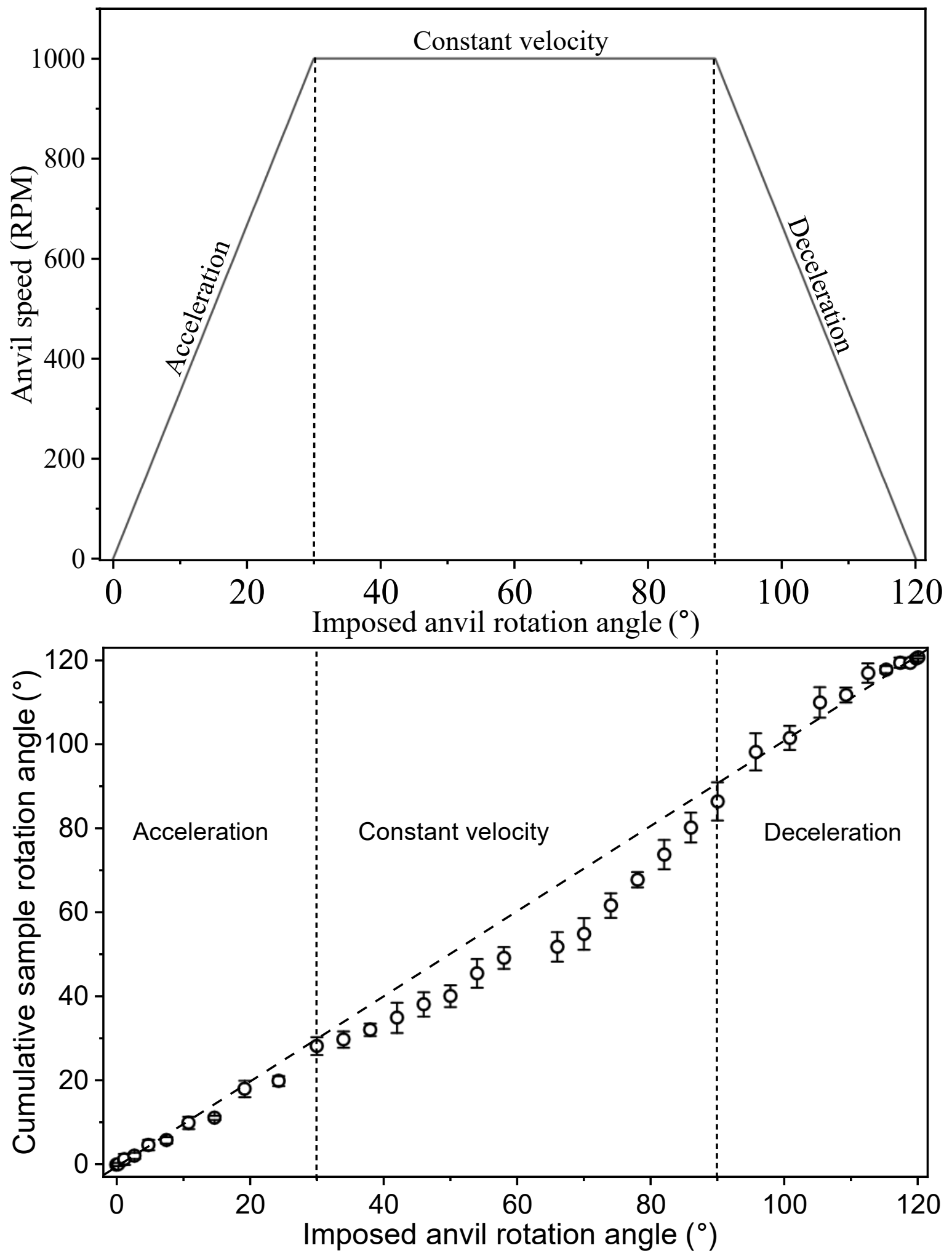

**Fig. S8: Results of the digital image correlation analysis of the Fe-7%Mn sample at 1,000 RPM and 4.3 GPa at *r* = 0 µm within dRDAC using a high-speed Photron camera (Run 3).** (a) Acceleration-constant velocity-deceleration profile with a constant speed portion of 1,000 RPM. (b) Cumulative sample surface rotation angle versus imposed anvil rotation angle. The vertical lines at 30 and 60 degrees separate the acceleration, constant velocity, and deceleration regions. The dotted line at 45° corresponds to ideal adhesion between diamond and sample.

One of the principal questions in high-pressure torsion is whether slippage occurs at the sample-anvil interfaces, especially at high rotation rates. For smooth diamonds, the friction stress $\tau_f$ is two times smaller than the yield strength[23,24], $\tau_y$ and slippage reaches 80% of the rotation angle of diamonds[2]. Recent theoretical analysis of gripping conditions in high-pressure torsion by Beygelzimer *et al.* suggests that for $\tau_f<\tau_y$, slippage becomes more pronounced at higher rotation rates[3,4]. To eliminate slippage, we introduced rough diamond anvils with an asperity of ~400 nm (instead of the traditional 10 nm), as described in the Methods section. Unlike smooth diamonds, these large diamond asperities penetrate the sample surface and, in theory, prevent sliding.

To demonstrate the lack of slippage, we performed in situ optical imaging of the Fe-7%Mn sample surface under torsion at an anvil speed of 1,000 RPM in run 3, using a high-speed camera at the HPCAT, Advanced Photon Source. High-speed imaging of the sample surface from the rotating piston-anvil side was performed using a Photron FASTCAM SA3 camera at 1000 fps. The sample surface was magnified ~6.66 times using a combination of an achromatic objective (Thorlabs AC254-150, focal length, $f \approx 150$ mm) and an imaging lens (OptoSigma DLB50.8-1000, $f \approx 1$ m). A pellicle beam combiner (8% reflectivity) introduced broadband white light. We illuminated the sample with an incandescent lamp and captured real-time images on the piston side.

This experiment was the first step in the new experimental loading program, designed based on the results from runs 1 and 2, which we will analyze in a separate paper. We used Almax easy Lab rough anvils with a culet size of 300 μm and an asperity of 400 nm for this experiment. Since simultaneous XRD measurements were not possible during the recording of the movie using the Photron FASTCAM SA3 camera, we chose a rotation speed of 1,000 RPM (instead of the maximum speed of 1,500 RPM in the current paper) to decrease the total rotation angle and reduce the sample thickness for the more productive following loading program. A Fe-7%Mn sample was pre-indented to ~134 μm thick, laser-cut to a diameter of 160 μm, and placed inside a Re gasket to minimize thickness reduction during torsion and achieve more uniform pressure along the radius. Because Re is much stronger than the Fe-7%Mn alloy, diamond asperities should penetrate less into Re than into the sample, which may promote sliding between the diamond and gasket and potentially cause sample sliding. Also, because sliding is expected to be higher at low pressures, we applied torsion at pressures as low as 4.3 GPa at $r = 0$ μm to provide more favorable conditions for sliding. For Digital Image Correlation (DIC) analysis, we extracted images from the video at a reduced frame rate of 10 fps (Supplementary Movie 1). We subjected the sample to torsion at the anvil speeds shown in Fig. S8a, with a constant-velocity portion of 1,000 RPM for 120° (30 ms).

It takes 10 ms (30°) to accelerate the Teknic motor to the maximum angular velocity of 1,000 RPM, then 10 ms (60°) at constant velocity, and finally 10 ms (30°) to decelerate to zero velocity. The optical image resolution was relatively poor because the longer piston length partially obstructs the light and the camera has a large pixel size. Supplementary Movie 1 provides a real-time movie of the sample surface motion.

The following equations for the rotation angle of an anvil were used:

$$\theta = 0.5\alpha_+ t^2 \text{ for } 0 \le t \le 10 \text{ ms}; \ \alpha_+ = 0.6^{\circ}/ms^2 \ \rightarrow \ \theta = 30^{\circ} \text{ at } t = 10 \text{ ms};$$

$$\theta = 30^{\circ} + \omega_0 \,(t - 10) \text{ for } 10 \le t \le 20 \, ms; \quad \omega_0 = 6^{\circ}/ms \rightarrow \ \theta = 90^{\circ} \text{ at } t = 20 \text{ ms};$$

$$\theta = 90^{\circ} - 0.5\alpha_- t^2 \text{ for } 20 \le t \le 30 \text{ ms}; \ \alpha_- = -0.6^{\circ}/ms^2 \ \rightarrow \ \theta = 120^{\circ} \text{ at } t = 30 \text{ ms}.$$

We performed DIC analysis using the open-source 2D subset-based digital image correlation software package Ncorr[5]. Using MATLAB, we extracted 11 images during the acceleration and deceleration stages and collected 15 images during the constant-velocity region with a 4° step size. When the images were paired with an angular separation above 5°, the DIC correlation failed because the subsets showed no correlation, so we analyzed each adjacent pair with the first image as the reference and the second as the current image. The 58-66° image pair was analyzed with an 8° step size because the DIC results for the 58-62° pair were unreliable. In Ncorr, the reference image is divided into smaller regions called subsets, and the displacement of the subsets is tracked for the deformed image by assuming homogeneous deformation within each subset[5]. The subset radius is 15 pixels, corresponding to a 31 × 31 pixel subset (1 pixel = 2.5526 µm), and we used a subset spacing of 5 pixels for all data pairs. As a result, neighboring subsets overlapped by 26 pixels (~84%), providing a smooth displacement field with better spatial resolution. Each subset is associated with a center point called the subset center, where the displacement is evaluated. We used the Gauss-Newton iteration, the norm of the difference vector, and the correlation coefficient to ensure convergence. We used a strain radius of 7 pixels for all data pairs. The strain radius is the radius of the neighborhood DIC subset centers, and a least-squares plane is fitted to these points to obtain the displacement gradients (the slope of the fitted surface). We determined the sample center just before the onset of sample rotation and kept it fixed throughout the analysis to ensure a common reference frame. We used a least-squares rigid-body fitting algorithm based on singular value decomposition[6] to calculate the rigid-body rotation between the reference and deformed image pairs. Because we obtained reliable DIC subset correlation only within a sample radius of 60 μm, we used data only within this radius for all analyses. We calculated the incremental sample rotation angle for each image pair from its corresponding rotation matrix and summed them to obtain the cumulative sample rotation angles. The uncertainty for each image pair was estimated from the distribution of the local rotation angles of all accepted DIC subsets using the median absolute deviation of the local rotation angles. The error bar for each image pair shows the corresponding cumulative sample rotation angle.

Fig. S8b shows the cumulative rotation angle of the sample extracted using DIC analysis versus imposed anvil rotation angle for the acceleration, constant velocity, and the deceleration stages. The plotted line under 45° corresponds to equal angles of rotation of an anvil and a sample, i.e., to complete adhesion and lack of relative sliding. The maximum sample rotation angle achieved is 120.7° with a deviation of 0.6%, which agrees well with the applied anvil rotation angle of 120°. Most calculated data points lie close to the adhesion line during acceleration and deceleration, indicating no sliding. In the constant velocity region, up to 60° of the imposed anvil rotation angles, the sample rotation angle deviates by less than 10°. Between 65° and 75°, a maximum deviation of up to 15° is observed; afterward, the deviation decreases. In the constant velocity region, rotation of the sample is visibly smaller than the calculated rotation of an anvil. However, this cannot be interpreted as sliding, because the anvil and sample rotation angles correspond well for the entire 10 ms at constant angular velocity. Assuming sliding at the beginning of the constant angular velocity stage, increments of sample rotations are greater than the increments of anvil rotations (slopes of both curves) at the end of the constant angular velocity stage, which is unphysical. The only noncontradictory way to explain the result is to assume that the motor imposes an anvil rotation speed of 1,000 RPM on average over the time interval $10 \leq t \leq 20\ ms$. But instantaneous values are initially slightly smaller and then are slightly larger than 1,000 RPM, and there is no sliding.

## V. Thickness calculation

We used a 1D X-ray absorption method to calculate sample thickness along various compression and shear $p$-$q$ paths. In the 16-ID-B beamline, a ~1 cm-long ion chamber is used to monitor X-ray counts entering the beamline hutch. The X-ray beam then passes through the focusing mirror assembly, the pinhole assembly, and finally the sample assembly before reaching the PILATUS3 X 2M CdTe detector. A PIN diode at the beam stop measures the transmitted intensity. The current from these counters is converted to a 0-5 V signal using a current preamplifier. This voltage is then fed to a voltage-to-frequency (V-to-f) converter, and a scaler counts it.

The sample thickness is calculated as follows.

The Beer-Lambert law[7] can be modified as

$$I_t = C I_0 e^{-\mu_m \rho t} \qquad \text{(S6)}$$

where $\mu_m$ is the mass attenuation coefficient of the material, $\rho$ is the density, and $t$ is the sample thickness. $I_0$ is the count at the ion chamber before the sample, and $I_t$ is the count measured at the PIN diode after the sample. We introduce $C$ as a scaling factor to account for the various losses (e.g., due to mirrors and windows) between the ID-B ion chamber, absorption correction of diamond anvils, and the PIN diode.

We have calculated $C$ using the initial and retrieved sample thicknesses and their 1D X-ray absorption coefficients. We use the C value calculated from the retrieved sample to estimate the sample thickness under higher-pressure, shear conditions, accounting for fluctuations in X-ray absorption counts in the ion chamber and the PIN diode.

The mass attenuation coefficient ($\mu$) of Fe-7%Mn is calculated from the mass attenuation coefficient of Fe (8.176 $cm^2/g$) and Mn (7.141 $cm^2/g$) at 30 keV[8], which is $\mu = (0.93\times8.176) + (0.07\times7.141) = 8.10355\ cm^2/g$.

The accuracy of the sample thickness calculation using this method arises from the error in calculating the mass attenuation coefficient (~ 0.1%), the signal fluctuations and noise in the ion chamber and PIN diode (~0.3- 0.5%), and the error in the calculation of densities of BCC and HCP phases using the formula[7]

$$\rho = \frac{Z\, M_{avg}}{N_A\, V} \qquad \text{(S7)}$$

where $M_{avg}$ is the average atomic mass of Fe-7%Mn, $N_A$ is Avogadro's number, Z is the formula unit, which is 2 for both BCC and HCP phases, and V is the volume of the BCC and HCP phases.

The error in the density calculation using the above approach is about $\leq 3\%$. Considering the above factors, the error in the thickness calculation in our approach is approximately 2 μm.

## VI. Lack of temperature rise during torsion in dRDAC

Since plastic straining with such high strain rates may cause significant heating, we evaluate here the maximum temperature in the sample using a simplified approach based on the finite element method (FEM). Eq. (S5) is used for calculating the plastic shear strain rate for the final sample thickness $h_f = 8.2\ \mu m$ at the end of torsion. Angular velocity $\omega$ follows the curve shown in Fig. S9, as in the experiment.

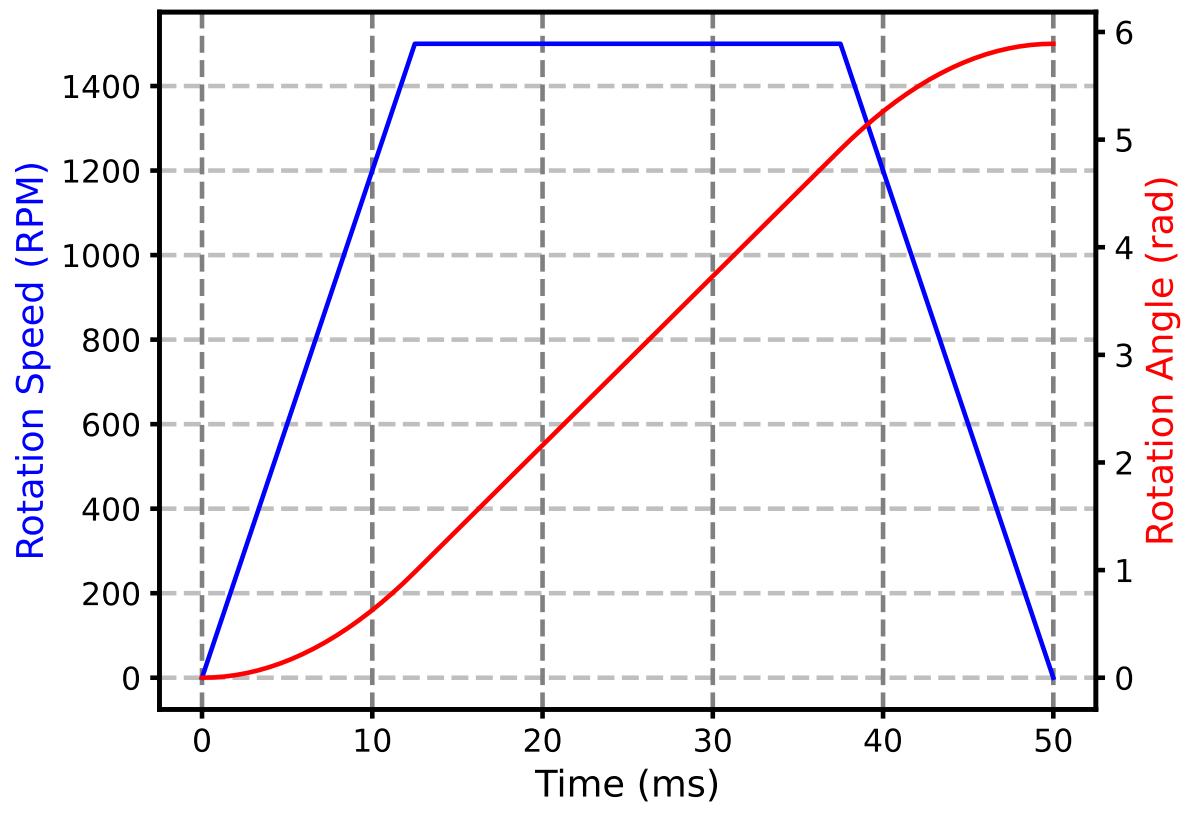

**Fig. S9: Rotation speed and angle of an anvil vs. time in the experiment and simulations.**

Heat generation rate due to plastic work per unit mass in the sample below the culet region ($r \leq 150$ μm) is calculated using the following equation

$$\dot{Q}_p = \eta \tau_y \dot{\gamma} \qquad \text{(S8)}$$

where the factor $\eta = 0.9$ considers that not the entire stress power dissipates into heat, and the portion $1 - \eta = 0.1$ of the stress power accumulates in a material as the energy of different defects; $\tau_y = \sigma_y/\sqrt{3}$ is the yield strength in shear according to the von Mises yield criterion. Outside the culet ($r > 150$ μm), heat production is very small because the FEM solution to the mechanical problem shows large sliding along the conical sample-anvil contact due to low friction and, consequently, a small plastic strain rate in that portion of the sample. Friction heat does not affect the solution under the culet because it mostly flows into the anvils and is in the major region far from the sample below the culet. Implementation is done in the FEM code Abaqus using the HETVAL subroutine.

| Material Properties | Sample (Fe) | Anvil (diamond) |
|---|---|---|
| Thermal conductivity, $k$ | $0.074\, W/(mm \cdot K)$[9] | $2.0\, W/(mm \cdot K)$[12] |
| Mass density, $\rho$ | $(7.874 \times 10^{-6}\, kg/mm^3)$[10] | $(3.515\, kg/mm^3)$[13] |
| Specific heat, $c_p$ | $464.4\, J/(kg \cdot K)$[9] | $547.01\, J/(kg \cdot K)$[14] |
| Yield strength in shear, $\tau_y$ | $1.1\, GPa$[11] | |
| Interface thermal properties | | |
| Convection coefficient (all free surfaces) | $1.0 \times 10^{-5}\, W/(mm^2 \cdot K)$ | |
| Sample-anvil interface contact conductance | $(200\, W/mm^2)$[15] | |

**Supplementary Table 14: Material properties used in thermal simulations.**

The resulting transient heat conduction equation is as follows:

$$\frac{\partial T}{\partial t} = \frac{k}{\rho c_p}\nabla^2 T + \frac{\dot{Q}_p}{\rho c_p} = \frac{k}{\rho c_p}\nabla^2 T + \frac{\eta \tau_y \omega r}{\rho c_p h} \qquad (S9)$$

For simplicity, we use pure Fe sample properties and present them in Supplementary Table 14.

Reference 15 uses the time-domain thermoreflectance (TDTR) technique to measure contact conductance between diamond and Pb, Au, Pd, Pt, and Al. Sample-anvil contact conductance is approximated to that of Pt because, out of all metals, Pt was the most like Fe based on electronic configuration and thermal conductivity values. In simulations, a change in contact conductance to 100 $W/mm^2$ showed no noticeable change in temperature distribution in the sample, indicating that the potential error in the value of the thermal conductance does not produce an essential error in the temperature.

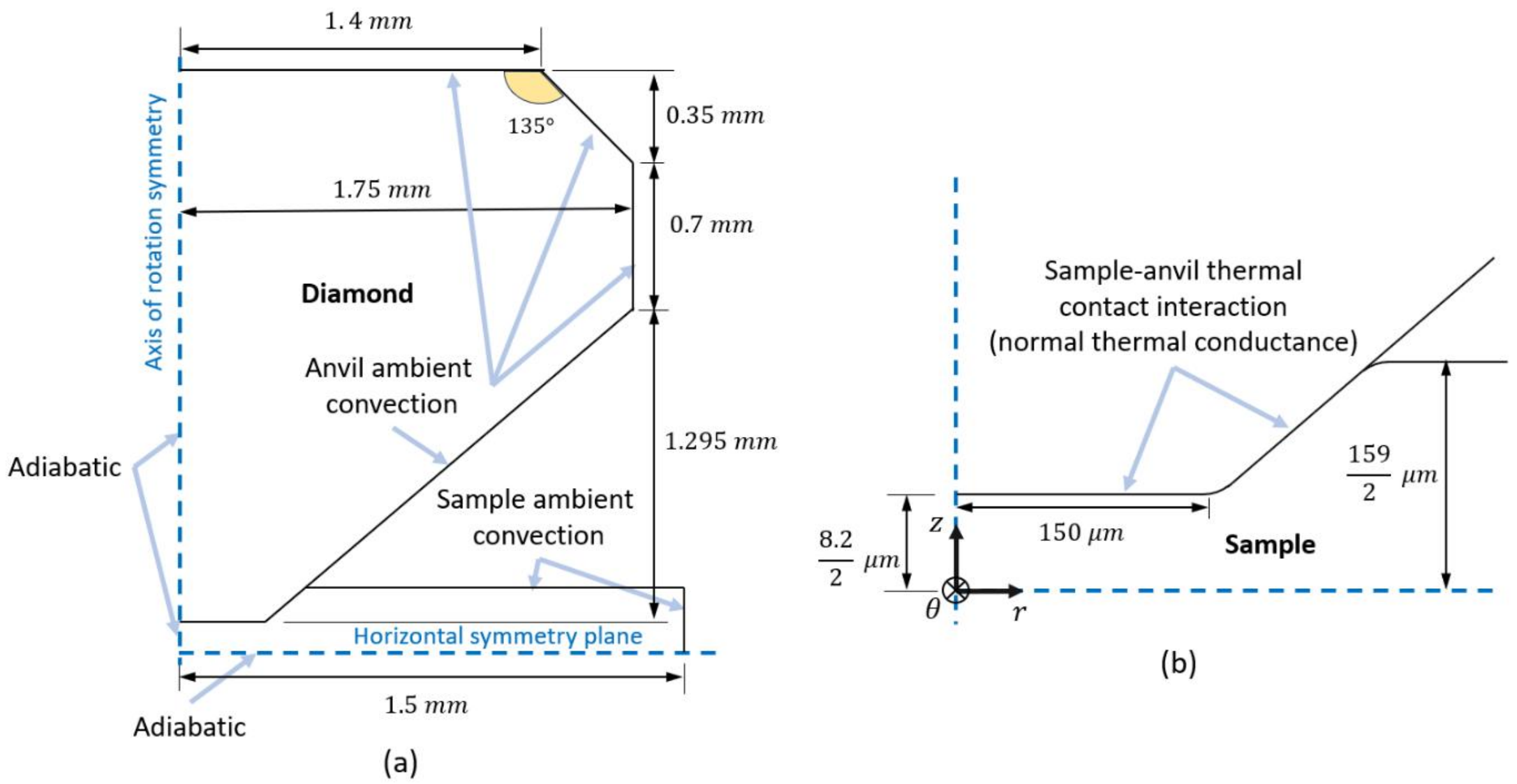

**Figure S10: Quarter-anvil and sample assembly geometry and thermal boundary conditions used for axisymmetric transient thermal simulation:** (a) the entire system and (b) the sample-anvil contact region.

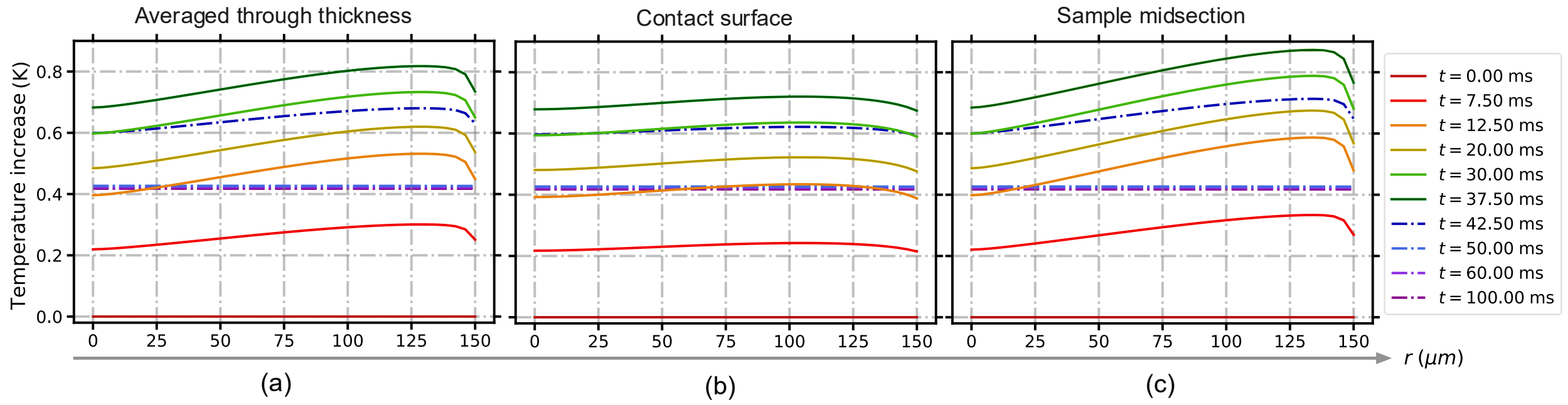

**Figure S11: Evolution of radial temperature distribution in the sample.** Solid lines represent increasing temperature, and dashed lines represent decreasing temperature. (a) Temperature averaged over the sample thickness. (b) Temperature at the sample-diamond contact surface. (c) Temperature at the horizontal symmetry plane.

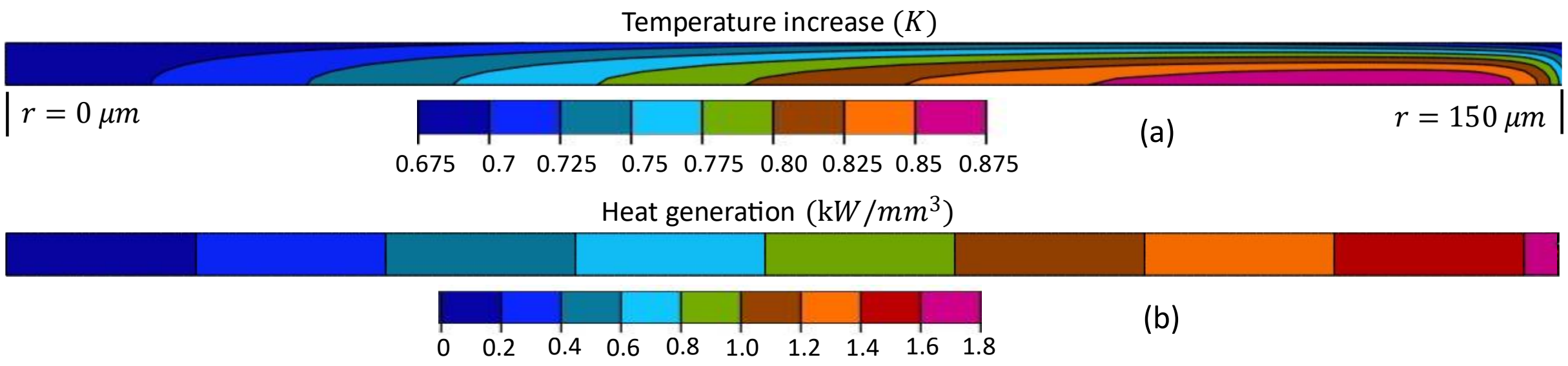

**Figure S12: Results of FEM temperature simulations.** (a) Distribution of the temperature increment with respect to the initial room temperature in the quarter of a sample section when the maximum temperature is reached at $t = 35\ ms$. (b) Input heat generation for 1,500 RPM.

The geometry of the entire system and the sample-anvil contact region are shown in Fig. S10.

Results of FEM simulations are presented in Figs. S11 and S12. The key point is that the maximum temperature increment in the sample at 1,500 RPM does not exceed 1 K, i.e., is negligible. Since the maximum temperature increment is proportional to $\tau_y \omega$, an increase of this product even by a factor of 10 or more should not affect PTs and microstructure evolution. This shows that dRDAC offers a unique way to completely decouple the effects of strain rate and temperature, and that a more precise coupled thermomechanical formulation and more precise determination of material properties should not change this conclusion.